\documentclass[%
 reprint,
 amsmath,amssymb,
 aps,
]{revtex4-2}

\usepackage{graphicx}
\usepackage{dcolumn}
\usepackage{bm}
\usepackage{xcolor}
\usepackage{amsmath}
\usepackage{isomath}
\usepackage{physics}
\usepackage{bbm}
\usepackage[bbgreekl]{mathbbol}
\usepackage{float}
\usepackage{soul}
\usepackage{upgreek}

\begin{document}

\preprint{APS/123-QED}

\title{Hierarchical Topological States without Dimension Reduction}

\author{Joel R. Pyfrom}
\affiliation{
	Department of Mechanical Engineering, University of Vermont, Burlington, VT 05405, USA}
\author{Kai Sun}
\affiliation{Department of Physics, University of Michigan, Ann Arbor, MI 48109, USA}
\author{Jihong A. Ma}
\email{Jihong.Ma@uvm.edu}
\affiliation{
Department of Mechanical Engineering, University of Vermont, Burlington, VT 05405, USA}
\affiliation{Department of Physics, University of Vermont, Burlington, VT 05405, USA}
\affiliation{Materials Science Program, University of Vermont, Burlington, VT 05405, USA}

\begin{abstract}
Topological insulators exhibit boundary states protected by bulk band topology, a principle first established in quantum systems and later extended to classical waves, including phononics. Conventionally, an $n$-dimensional bulk with nontrivial topology hosts $(n-1)$-dimensional topologically protected boundary states, which may be further gapped out by breaking the symmetry that protects them, potentially leading to the emergence of $(n-2)$-dimensional, or even lower-dimensional topological states, as in higher-order topological insulators. In this work, we introduce an alternative mechanism for gapping out topological states and forming new topological modes within the resulting gap without further unit-cell symmetry breaking or dimension reduction. Using one- and two-dimensional Su-Schrieffer-Heeger (SSH) models, we show that controlled repositioning of topological domain walls enables the construction of hierarchical unit cells that gap out the original domain-wall states while preserving the underlying symmetry. This process produces higher-hierarchical-level topological states, characterized by a generalized winding number, and can be iterated to realize multiple - potentially infinite - hierarchical levels of topological states. Our approach expands the conventional topological classification and offers a versatile route for engineering complex networks of protected modes in higher dimensions.

\end{abstract}

\maketitle

\section{\label{sec:level1} Introduction}

Topological states of matter exhibit a range of intriguing anomalous effects that transcend the conventional paradigm of symmetry breaking~\cite{thouless1982quantized,haldane1988model,hasan2010colloquium,qi2011topological}. These states are characterized by nontrivial bulk band topology, often quantified by topological invariants, and manifest through the bulk-boundary correspondence: an insulating bulk coexists with discrete boundary modes residing in the bandgap. Unlike trivial boundary states induced by lattice termination, topological boundary states are protected against backscattering from defects, disorder, and sharp corners - a property known as topological protection. This robustness underpins potential applications in low-dissipation transport and wave control~\cite{chang2015zero,ma2023phonon,linder2010unconventional}.

Recent advances have extended these concepts to classical platforms such as acoustics, phononics, and mechanics, enabling realizations of analogs to the quantum Hall effect~\cite{nash2015topological,wang2015topological}, quantum spin Hall effect (QSHE)~\cite{mousavi2015topologically,he2016acoustic}, quantum valley Hall effect~\cite{pal2017edge,lu2017observation,ma2019valley,liu2018tunable}, and toplogical polarization observed in Maxwell lattices~\cite{ma2018edge,mao2018maxwell,sun2012surface,kane2014topological,paulose2015topological}. While these states are often gapless, further unit-cell symmetry breaking can open gaps in their edge spectra, enabling higher-order topological states (HOTSs) within these gaps, however with dimension reduction. For example, intersecting two gapped one-dimensional (1D) topological edge states with distinct gap topologies in a 2D QSHE system can yield second-order corner modes~\cite{zhang2019second,chen2021corner}. In general, HOTSs appear in dimensions one lower than the parent topological phase due to the further unit-cell symmetry reduction, requiring at least a 2D lattice for second-order states~\cite{cualuguaru2019higher}, with the maximal order in 3D systems being three~\cite{weiner2020demonstration}.

In this work, we introduce an alternative hierarchical mechanism to gap out topological states and enable the emergence of new topological modes within the induced gap. Using a mechanical analog of the Su-Schrieffer-Heeger (SSH) model~\cite{su1979solitons,su1980soliton} as a prototype, we show that original topological domain-wall states (TDWSs) formed at the interface between first-hierachical-level (1st-HL) unit cells of distinct phases can be gapped out without further breaking the underlying 1st-HL symmetry. Consequently, no dimensional reduction is required. This is achieved by strategically positioning two adjacent domain walls to construct a second-hierarchical-level (2nd-HL) unit cell, which opens a new gap between the originally degenerate TDWSs. When topologically nontrivial, this gap hosts a 2nd-HL topological state. Iterating this process - nesting dual 2nd-HL domain walls into a third-hierarchical-level (3rd-HL) unit cell - yields 3rd-HL states, and, in principle, the hierarchy can extend to arbitrary levels and higher dimensions. The number of hierarchical states is predicted by a generalized winding number and corroborated by a $\mathbb{Z}_2$  invariant, with finite-lattice simulations confirming bulk-boundary correspondence. Furthermore, our framework determines the number of TDWSs at interfaces between phases of different hierarchical levels. 

Although demonstrated in a classical mechanical SSH model, the theory is fundamentally system-agnostic: its mathematical structure mirrors that of quantum tight-binding models, ensuring that our predictions - including winding numbers and hierarchical bulk-boundary correspondence - apply equally to quantum condensed-matter systems.

\section{\label{sec:level1} Results and Discussion}

\subsection{One-Dimensional Hierarchical Su-Schrieffer-Heeger Lattices}
\subsubsection{Generalized Winding Number Calculation for One-Dimensional Lattices}

Before addressing higher-HL topology, let's first briefly review the determination of topological phases using a 1D SSH chain consisting of identical masses $m$ and alternating spring constants $c_1$ and $c_2$ (see Fig.~\ref{fig:S1} in the Supplementary Material). Although the phonon dispersions for $c_1>c_2$ and $c_1<c_2$ appear identical, the latter exhibits a parity inversion of the unit-cell eigenmode at the Dirac point $k=\pi/a$, where $k$ is the wave number and $a$ the lattice constant. This inversion signals a topologically non-trivial phase, whereas the former configuration is trivial. 

The topological distinction is quantified by the winding number $W$, computed from the stiffness matrix $\bm{H}(k)$~\cite{chiu2016classification} (the expression of $\bm{H}(k)$ is presented in Eqn.~\ref{eq:Hmatrix} of the Supplementary Material):
\begin{eqnarray} 
	W = \int^{\pi/a}_{-\pi/a} \frac{1}{4 \pi i} tr(\bm{\sigma}_3 \bm{H}'^{-1}\partial_{k}\bm{H}') dk,
	\label{winding_number_original}
\end{eqnarray} 
where $\bm{\sigma}_3$ is the third Pauli matrix and $\bm{H}'=\bm{H}(k) - (c_1 + c_2)\bm{I}$ (where $\bm{I}$ represents the identity matrix) is a chiral-symmetric form obtained by subtracting the diagonal elements of $\bm{H}(k)$. Since such a subtraction only shifts the eigenvalue spectrum down by a constant $\omega_0^2=(c_1+c_2)/m$ without altering the eigenvectors, allowing normalized eigenvalues symmetric about $\omega^2/\omega_0^2=0$. This symmetric representation preserves the intrinsic chiral symmetry of the SSH model, ensuring an exact mapping to tight-binding models (see Section I of the Supplementary Material). Consequently, all topological invariants and boundary states discussed here translate directly to quantum SSH systems.

Alternatively, the winding number $W$ can be inferred from the $\mathbb{Z}_2$ invariant via eigenvector parity~\cite{fu2007topological}. For each time-reversal-invariant momentum $\Gamma_i$ ($ka=0,\pi$), we compute 
\begin{eqnarray} 
	\delta_i = \prod_{\nu=1}^{N} \zeta_{\nu}(\Gamma_i),
	\label{eq:Z2_sigma}
\end{eqnarray} 
where $\zeta_\nu$ denotes the parity of the $\nu$-th band eigenvector and $N$ is half the number of bands. The invariant is then 
\begin{equation}
		(-1)^W = \prod_{i} \delta_{i}.
		\label{eq:nu}
\end{equation}
yielding $W = 1$ for a non-trivial phase and $W=0$ for a trivial phase. 

For the standard SSH model, both methods yield equivalent results and are often used interchangeably. However, the $\mathbb{Z}_2$ index captures only the parity of the winding number, $(-1)^W=\mathbb{Z}_2$, and thus cannot distinguish phases with $W>1$ or systems lacking inversion symmetry. In contrast, the generalized winding number provides the full integer $W$ and remains applicable to more complex hierarchical models. Although this approach typically requires integration over the entire Brillouin zone, we derive a closed-form expression that depends solely on material ($c_1$, $c_2$) and structural ($d_n$, $D_n$) parameters, bypassing eigenstate calculations and enabling rapid topology determination from structural data alone. Both methods are presented here, and their agreement confirms the robustness of our generalized $W$ characterization. Importantly, the generalized $W$ is essential for higher-hierarchical-level states, particularly in systems with long-range interactions, where $W$ can exceed unity, as demonstrated in our previous study~\cite{rajabpoor2023breakdown}.

Combining two lattice phases with different $W$'s creates a domain boundary hosting a number of TDWSs equal to $\Delta W$. This is illustrated by constructing a supercell with a central domain boundary and applying Bloch boundary conditions at the two matching ends (Fig.~\ref{fig:S2}, Supplementary Material). The resulting dispersion exhibits a single zero-frequency TDWS within the bulk bandgap $\Delta \omega_B^2$, consistent with $\Delta W=1$. Moreover, $W$ predicts the number of topological edge states (TPESs) in finite lattices with proper terminations~\cite{rajabpoor2024gap}, exemplifying bulk-boundary correspondence - though this correspondence may fail in the presence of long-range interactions~\cite{rajabpoor2023breakdown}.

\begin{figure*}
	\centering
	\includegraphics[scale=0.55]{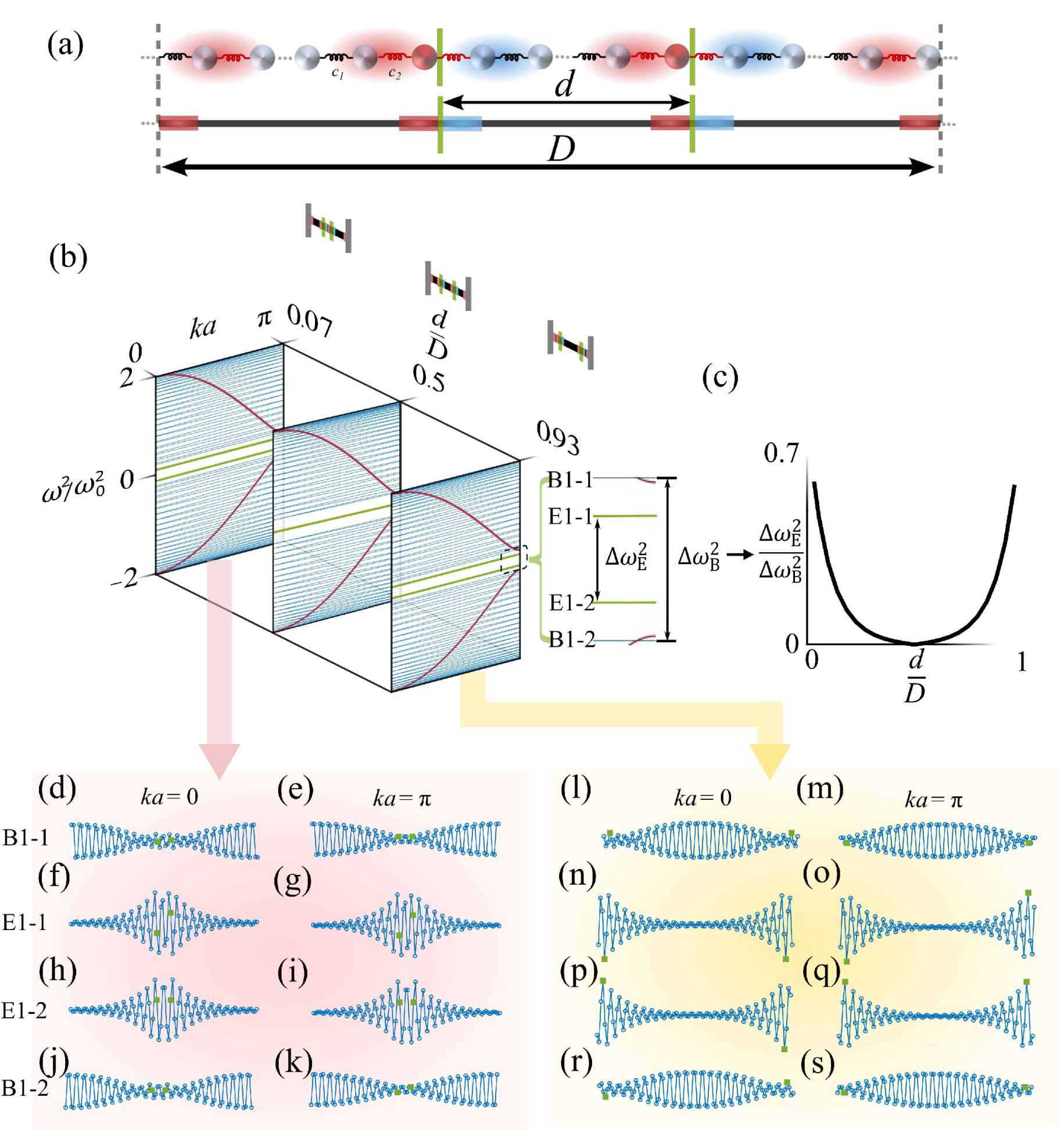}
	\caption{One-dimensional hierarchical lattice and its topological properties. (a) Schematic of a second-hierarchical-level (2nd-HL) unit cell comprising 96 identical masses connected by alternating springs of stiffness $c_1$ (black) and $c_2$ (red). Two first-hierarchical-level (1st-HL) domain walls are marked by vertical green bars. Red and blue shading highlight distinct 1st-HL arrangements near domain walls and terminations. The 2nd-HL cell, bounded by gray dashed lines, spans two domain walls; $d$ and $D$ denote the number of mass pairs between the domain walls and within the full 2nd-HL cell, respectively. (b) Phonon dispersion for $c_1>c_2$ showing bulk bands (blue), 1st-HL topological domain-wall states (TDWSs, green), and 1st-HL unit cell dispersions (red) for various $\frac{d}{D}$ ratios. (c) Normalized 2nd-HL bandgap $\Delta \omega_E^2/\Delta \omega_B^2$ between TDWSs (E1-1/2) and bulk bands (B1-1/2) as a function of $\frac{d}{D}$. (d-s) Mode shapes at (d,f,h,j,l,n,p,r) $ka=0$ and (e,g,i,k,m,o,q,s) $\pi$ for bands B1-1/2 and E1-1/2, comparing configurations with (d-k) $\frac{d}{D}=0.07$ and (l-s) $\frac{d}{D}=0.93$.}
	\label{fig:Interface_distance}
\end{figure*}

Doubling the supercell introduces two domain walls, and thus two degenerate zero-frequency TDWSs [Fig.~\ref{fig:Interface_distance}(a)]. Defining $d$ as the number of unit cells ($i.e.$, mass pairs) between domain walls and $D$ as the total mass pairs in the doubled supercell, the ratio $\frac{d}{D}=\frac{1}{2}$ yields exact degeneracy at $\omega^2/\omega_0^2=0$ [Fig.~\ref{fig:Interface_distance}(b)]. Deviating from this ratio splits the degeneracy and opens a secondary bandgap $\Delta \omega_E^2$, whose normalized size $\Delta \omega_E^2/\Delta \omega_B^2$ grows as $\frac{d}{D}$ increases [Fig.~\ref{fig:Interface_distance}(b,c)]. The critical transition occurring at $\frac{d}{D}=\frac{1}{2}$ arises from a competition between the intra- and inter-supercell domain-wall coupling, though additional factors influence the gap topology, as discussed later.

A key question is whether this emergent bandgap between the two TDWSs possesses topological character. A quick examination of the supercell mode shapes at $ka=0$ and $\pi$ reveals that, for $c_1>c_2$ and the configuration in Fig.~\ref{fig:Interface_distance}(a), parity inversion between the two TDWSs occurs only when $\frac{d}{D}>\frac{1}{2}$. This behavior implies a potential topological transition, where the bandgap may be topologically non-trivial for $\frac{d}{D}>\frac{1}{2}$ and trivial otherwise. For clarity, in our following discussion, we define the second-hierarchical-level (2nd-HL) unit cell as a supercell containing two domain walls, while the first-hierarchical-level (1st-HL) unit cell refers to the basic SSH element of two masses connected by $c_1$ and $c_2$.

At this stage, the ratio $\frac{d}{D}$ emerges as a key parameter governing the topology of the 2nd-HL unit cell. However, careful inspection of Fig.~\ref{fig:Interface_distance}(a) suggests that the 2nd-HL topology may also depend on the local spring configurations at Bloch boundaries and domain walls - particularly the topological properties of the underlying 1st-HL unit cells. To systematically explore these dependencies, we compute topological invariants using the framework established in Eqn.~\ref{winding_number_original}, supplemented by the $\mathbb{Z}_2$ invariant formalism from Eqns.~\ref{eq:Z2_sigma} and \ref{eq:nu}.

\begin{figure}
	\centering
	\includegraphics[scale=0.41]{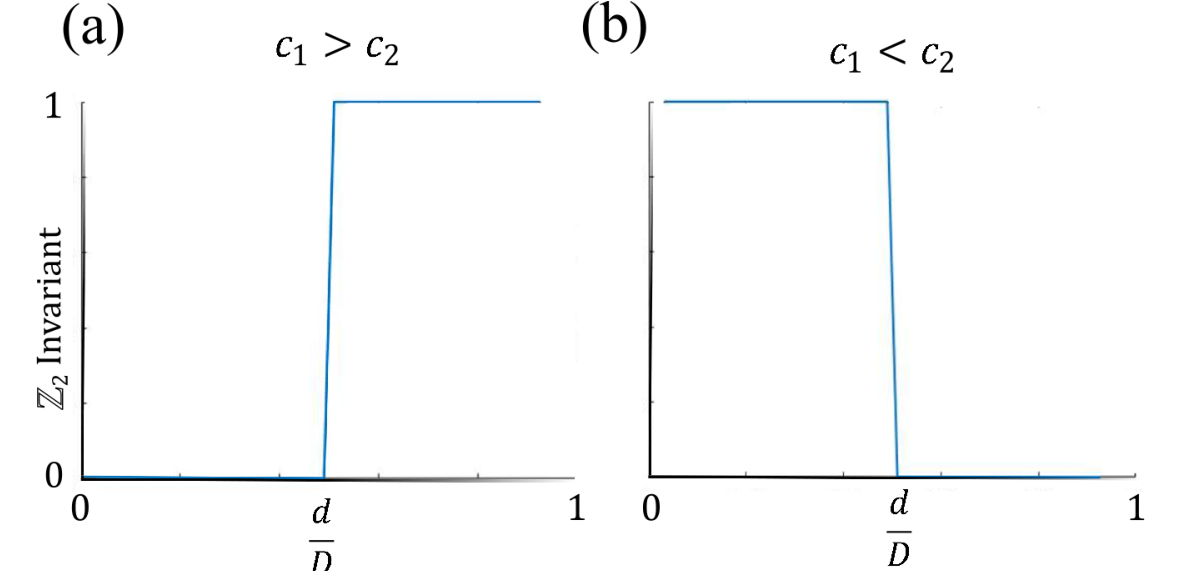} 
	\caption{Evolution of the $\mathbb{Z}_2$ invariant with respect to $\frac{d}{D}$ when (a) $c_1>c_2$ and (b) $c_1<c_2$ at the boundaries of the second-order unit cell.}
	\label{fig:z2}
\end{figure}

Notably, the $\mathbb{Z}_2$ invariant can be directly applied to the 2nd-HL unit cell, as its evaluation relies only on eigenvectors. As shown in Fig.~\ref{fig:z2}, the $\mathbb{Z}_2$ invariant depends critically on both the ratio $\frac{d}{D}$ and the strengths of $c_1$ and $c_2$, $i.e.$, the 1st-HL topology at periodic boundaries. Importantly, the invariant shows no dependence on the 1st-HL unit cell topology localized at the two domain walls themselves. 

The direct application of Eqn.~\ref{winding_number_original} to compute the winding number $W$, however, is not as straightforward. This complexity arises because the matrix $\bm{\sigma}_3$ is a $2\times2$ Pauli matrix associated with a diatomic 1st-HL unit cell. It needs to be scaled up to an equivalent $2N \times 2N$ matrix $\bm{\Theta}$ for a system with $N$ mass pairs ($N=D$ for the 2nd-HL unit cell). Since $\bm{\sigma}_3$ is Hermitian, unitary ($\bm {\sigma}_3^2 = \bm{I}$) and anti-commutes with the stiffness matrix ($\bm{\sigma}_3^{-1} \bm{H} \bm{\sigma}_3 = -\bm{H}$), these essential algebraic properties must be preserved in the enlarged representation. This is achieved by defining 
\begin{equation}
	\bm{\Theta}=\bm{I}_{2N\times 2N} \otimes \bm{\sigma}_3 = 
	\begin{bmatrix}
	1 & 0 &...& 0 & 0   \\
	0 & -1 && 0 & 0  \\
	&\vdots&\ddots&\vdots&\\
	0 & 0 && 1  & 0  \\
	0 & 0 &...& 0 & -1  \\
	\end{bmatrix}_{2N\times 2N}.
	\label{eq:large_sigma3}
\end{equation} 
where $\otimes$ denotes the Kronecker product~\cite{shi2021disorder,maffei2018topological,manmana2012topological}.

The winding number then generalizes to
\begin{eqnarray} 
	W = \int^{\pi/a}_{-\pi/a} \frac{1}{4 \pi i} tr(\bm{\Theta} \bm{H}'^{-1}\partial_{k}\bm{H}') dk,
	\label{eq:gen_W}
\end{eqnarray} 
where $\bm{H}'$ is the chiral Hamiltonian of the 2nd-HL unit cell. 

A detailed derivation (see Section II of the Supplementary Material) yields the compact form:
\begin{equation}
	W=
\begin{cases}
	0  \quad\text{for} \, &\left(\frac{c_1}{c_2}\right)^{D-2d}>1\\
	1  \quad\text{for} \, &\left(\frac{c_1}{c_2}\right)^{D-2d}<1
\end{cases}
\label{eq:gen_W_express}
\end{equation}
where $c_1$ and $c_2$ are the stiffnesses at the periodic boundaries of the 2nd-HL unit cell. 

Thus, the topology of the 2nd-HL unit cell depends on (i) the boundary topology of the 1st-HL cells and (ii) the ratio $\frac{d}{D}$, where $d$ is the number of 1st-HL cells between domain walls and $D$ is the total number of mass pairs in the 2nd-HL unit cell. The resulting winding number $W$ can be classified into four distinct cases: 
\begin{itemize}
	\item \textbf{Type I:} $\frac{d}{D}>\frac{1}{2}$ and $c_1>c_2$ at the boundaries $\Rightarrow$ $W=1$.
	\item \textbf{Type II:} $\frac{d}{D}<\frac{1}{2}$ and $c_1>c_2$ at the boundaries $\Rightarrow$ $W=0$.
	\item \textbf{Type III:} $\frac{d}{D}>\frac{1}{2}$ and $c_1<c_2$ at the boundaries $\Rightarrow$ $W=0$.
	\item \textbf{Type IV:} $\frac{d}{D}<\frac{1}{2}$ and $c_1<c_2$ at the boundaries $\Rightarrow$ $W=1$.
\end{itemize}

\begin{figure}[H]
	\centering
	\includegraphics[scale=0.28]{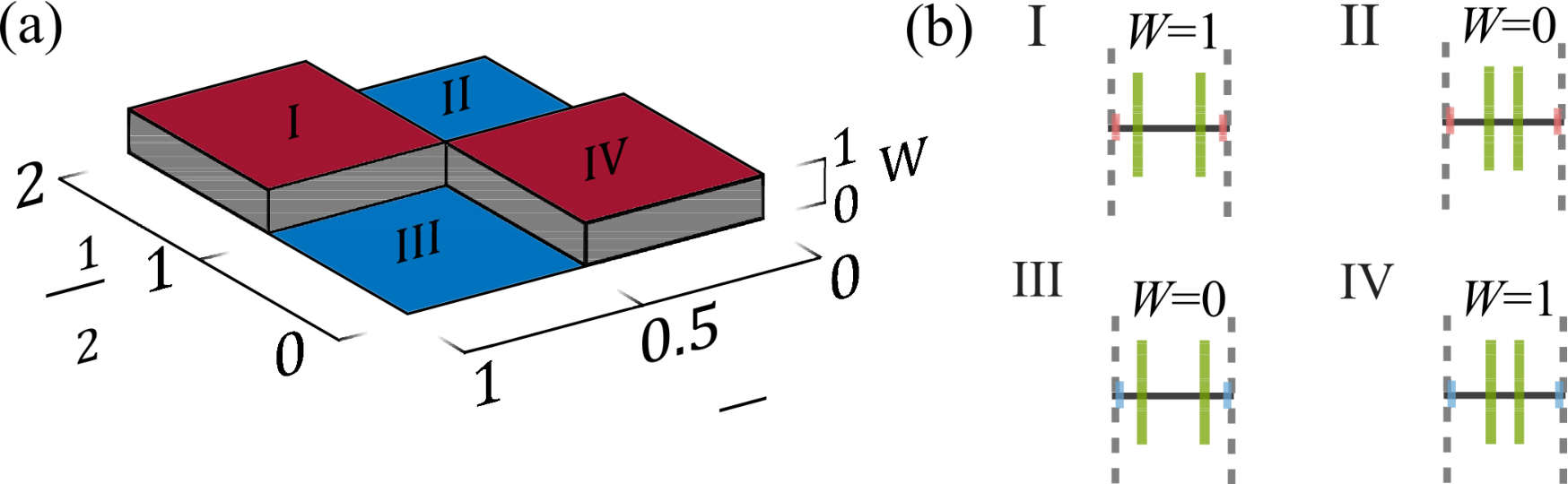} 
	\caption{Topological characterization of second-hierarchical-level (2nd-HL) unit cells. (a)2nd-HL winding number $W$ as a function of the spring constant ratio $\frac{c_1}{c_2}$ and structural ratio $\frac{d}{D}$, assuming $c_1$ precedes $c_2$ in the lattice sequence. (b) Schematic representations of the four 2nd-HL unit cell types labeled in (a), using the simplified notation of Fig.~\ref{fig:Interface_distance}(a). Boundary colors indicate spring configurations: red for strong-spring-start ($c_1>c_2$) and blue for weak-spring-start ($c_1<c_2$). All cases have $D=37$ mass pairs, the number of 1st-HL unit cells between domain walls (green bars), $d$, being 34 for Type I, 2 for Type II, 33 for Type III, and 3 for Type IV.}
	\label{fig:W3D}
\end{figure}

 Fig.~\ref{fig:W3D} illustrates the dependence of $W$ on the spring constant ratio $\frac{c_1}{c_2}$ and the structural ratio $\frac{d}{D}$, along with schematic representations of the corresponding 2nd-HL unit cells.

By iteratively assembling $n$th-hierarchical-level (HL) unit cells from pairs of $(n-1)$th-level domain walls at prescribed separations, the corresponding TDWSs are successively gapped (Fig.~\ref{fig:3rd_HL} in the Supplementary Material). This iterative process exhibits a self-similar pattern across hierarchical levels and bears a striking resemblance to fractal-like structures in reciprocal space when $n$ becomes large. The generalized winding number for the nnnth-HL unit cell is:
\begin{equation}
	W_n=
    \begin{cases}
	1 \quad \text{for} \, & \abs{A}_n<1 \\
	0 \quad \text{for} \, & \abs{A}_n>1,
\end{cases}
\label{eq:Wn}
\end{equation}
where $\abs{A}_n$ is a conditioning coefficient determining the topological state characterized by the $n$th-HL winding number $W_n$. $\abs{A}_n$ is recursively defined by the conditioning coefficients from the $(n-1)$th-HL unit cells located between ($\abs{A}_{n-1}^{in}$) and outside ($\abs{A}_{n-1}^{out}$) the $(n-1)$th-HL domain walls, along with the structural parameters $d_{n}$ and $D_{n}$. Here, $d_{n}$ denotes the number of $(n-1)$th-HL unit cells between the domain walls, and $D_n$ is the total number of such unit cells in the $n$th-HL cell. Specifically,
\begin{eqnarray}
    \abs{A}_n=\left(\abs{A}_{n-1}^{in}\right)^{d_n}\left(\abs{A}_{n-1}^{out}\right)^{D_n-d_n}.
    \label{eq:An}
\end{eqnarray}

Within our 1D SSH framework, the 1st-HL conditioning coefficients are $\abs{A}_1^{out}=\frac{c_1}{c_2}$ and $\abs{A}_1^{in}=\frac{c_2}{c_1}$. Substituting these into the recursive relation yields the 2nd-HL conditioning coefficient: 
\begin{eqnarray}
	\abs{A}_2=\left(\frac{c_2}{c_1}\right)^{d_2}\left(\frac{c_1}{c_2}\right)^{D_2-d_2}\\
    =\left(\frac{c_1}{c_2}\right)^{D_2-2d_2}.
	\label{A2_re}
\end{eqnarray}
A complete derivation of the general expressions for $\abs{A}_n$ and $W_n$, together with their implications for predicting higher-HL topological states, is provided in Section III of the Supplementary Material.

The following subsections discuss the influence of 2nd-HL winding numbers on edge states in finite 2nd-HL lattices, as well as the TDWSs generated by both different 2nd-HL cell configurations and mixtures of hierarchical levels.

\subsubsection{Second-Hierarchical Topological Edge States of a Finite Lattice}

To validate the number of 2nd-HL topologically protected edge states (TPESs) predicted by the 2nd-HL winding numbers in Eqn.~\ref{eq:gen_W}, we constructed four types of finite lattices from 2nd-HL cells with boundary terminations chosen to host zero-frequency ($\omega^2/\omega_0^2=0$) edge modes when permitted. As established in our previous work~\cite{rajabpoor2024gap}, such TPESs require identical diagonal entries in the global stiffness matrix, achieved by attaching grounding springs to all terminal and interfacial masses so that each mass retains an effective stiffness of $c_1 + c_2$ via its connection. This termination scheme is adopted throughout. 

\begin{figure*}
	\centering
	\includegraphics[scale=0.44]{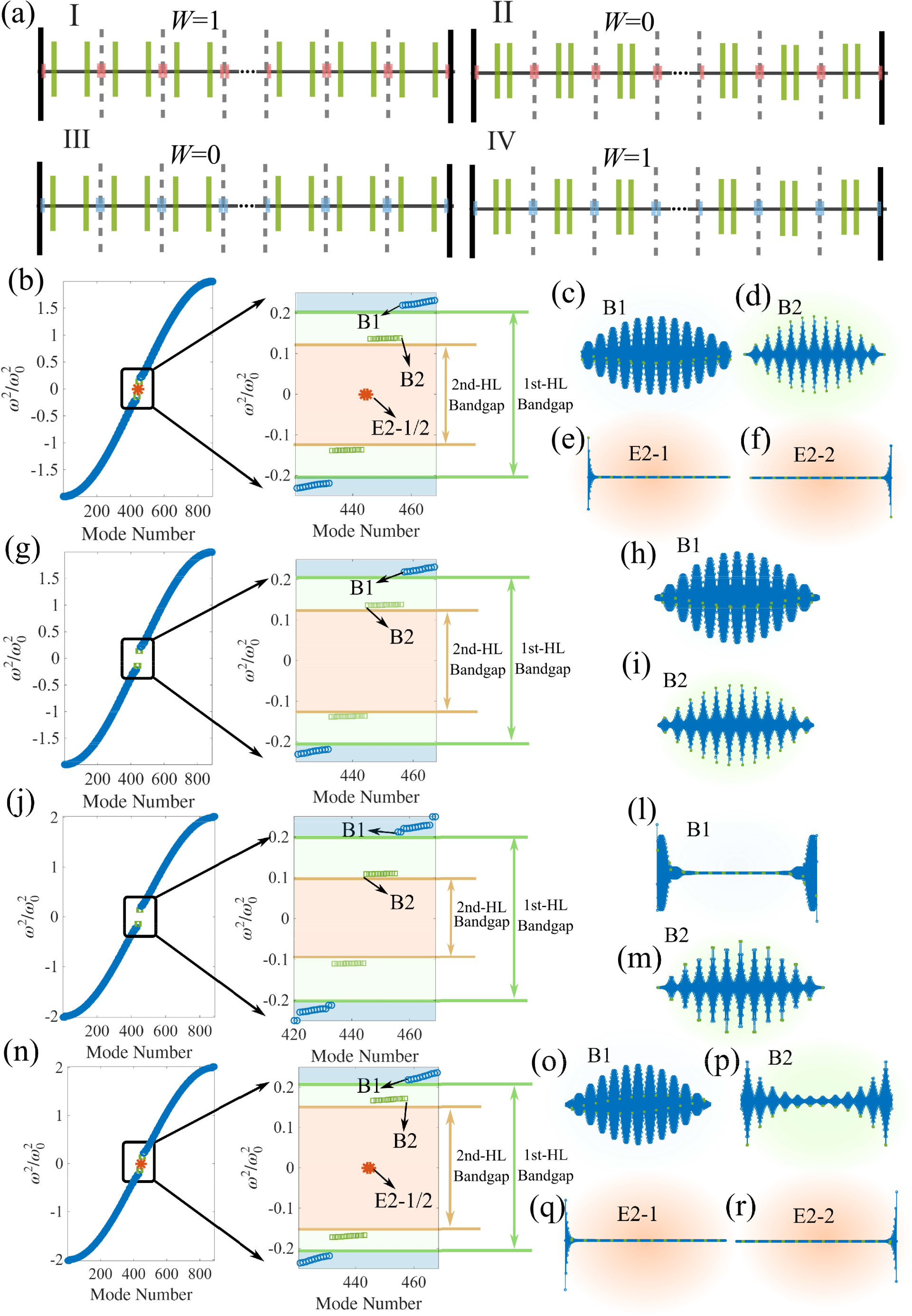} 
	\caption{Spectral and modal characteristics of one-dimensional finite hierarchical lattices. (a) Schematics of four finite lattices, each containing 12 second-hierarchical-level (2nd-HL) unit cells from Fig.~\ref{fig:W3D} (b). Black bars mark boundary masses with grounding springs to maintain an effective stiffness of $c_1+c_2$ for all masses. (b,g,j,n) Normalized eigenfrequencies ($\omega^2/\omega_0^2$) for lattices I-IV in (a), with zoomed-in views (right panels) highlighting 1st-HL bulk modes (B1, blue), 1st-HL bandgaps (green, hosting 2nd-HL bulk modes, B2), and 2nd-HL bandgaps (orange, potentially hosting 2nd-HL edge states, E2-1/2). (c-f,h-i,l-m,o-r) Representative mode shapes of B1, B2, and E2-1/2 (if present) modes for each lattice. Green squares denote 1st-HL domain wall masses.}
	\label{fig:fixed_boundary}
\end{figure*} 

The four finite lattices are presented in Fig.~\ref{fig:fixed_boundary} (a), each realizing one of the 2nd-HL cell types in Fig.~\ref{fig:W3D} (b). Eigenvalue analyses reveal consistent 1st- and 2nd-HL bulk gaps across all lattice types [Fig.~\ref{fig:fixed_boundary} (b,g,j,n)]. End-localized TPESs at $\omega^2/\omega_0^2$ appear only for Types I [Fig.~\ref{fig:fixed_boundary} (e,f)] and IV [Fig.~\ref{fig:fixed_boundary} (q,r)] and are absent in Types II and III. This finding aligns well with the results of their respective winding number calculations.

Notably, Types I and IV terminate with opposite boundary configurations: Type I with $c_1>c_2$, while Type IV with $c_1<c_2$. Our earlier study of the standard SSH chains (1st-HL only) revealed that TPESs at $\omega^2/\omega_0^2$ emerge exclusively for $c_1<c_2$ terminations, whereas $c_1>c_2$ exhibits no TPESs~\cite{rajabpoor2024gap}. Thus, the presence (absence) of TPESs for Type I (Type III) defies expectations based solely on boundary truncation and affirms their topological origin: the number of edge states is determined by the 2nd-HL winding number, rather than being merely an artifact of boundary truncation effects. In practice, when the finite lattice is composed of 2nd-HL unit cells, the interplay between the terminating 1st-HL cells and the normalized TDWS separations determines the 2nd-HL winding number, which in turn dictates the presence of TPESs.

\subsubsection{Second-Hierarchical Topological Domain-Wall States}

To verify the number of 2nd-HL TDWSs predicted by the 2nd-HL winding numbers (Eqn.~\ref{eq:gen_W}), we examine three distinct domain wall scenarios:
\begin{itemize}
\item \textbf{Between Type I ($W=1$) and Type II ($W=0$)} unit cells.
\item \textbf{Between Type III ($W=0$) and Type I ($W=1$)}.
\item \textbf{Between Type I ($W=1$) and Type IV ($W=1$)}.
\end{itemize}

\begin{figure*}
	\centering
	\includegraphics[scale=0.48]{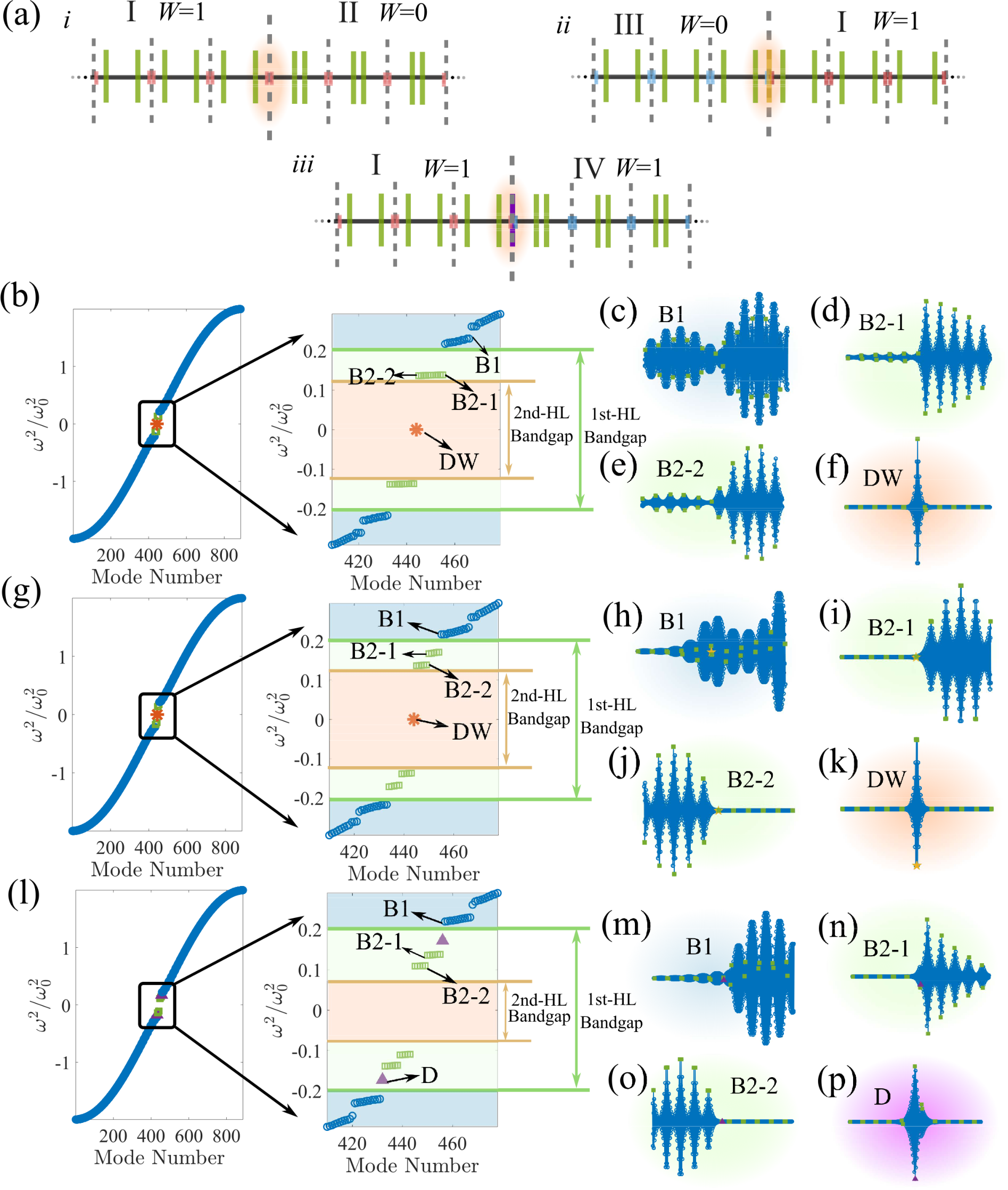} 
	\caption{Spectral and modal characteristics of one-dimensional hierarchical domain wall states. (a) Schematics of three finite lattices featuring 2nd-HL domain walls (orange), each with six 2nd-HL per side. (b,g,l) Normalized eigenfrequencies ($\omega^2/\omega_0^2$) for Scenarios $\it{i}$-$\it{iii}$ in (a), with zoomed-in views (right panels) highlighting: 1st-HL bulk modes (B1, blue), 1st-HL bandgaps (green, hosting 2nd-HL bulk modes, B2), and 2nd-HL bandgaps (orange, potentially hosting 2nd-HL domain-wall states, DW). (c-e,h-j,m-o) Representative mode shapes of B1, B2-1/2, and DW (if present). Green squares, yellow stars, and purple triangles mark 1st-HL domain walls, 2nd-HL topological domain walls, and 2nd-HL trivial domain walls, respectively.}
	\label{fig:interfacing_plot}
\end{figure*} 

The corresponding lattices, spectra, and mode shapes are shown in Fig.~\ref{fig:interfacing_plot}. In all cases, 2nd-HL bandgaps arise because $\frac{d}{D}\neq \frac{1}{2}$ on both sides of the 2nd-HL domain walls. These gaps are bounded by 2nd-HL bulk modes that involve simultaneous excitation of multiple 1st-HL TDWSs [Fig.~\ref{fig:interfacing_plot} (d,e,i,j,n,o)], enabling higher-HL topological states to form within them.

Scenarios $\it{i}$ and $\it{ii}$ exhibit a winding number difference $\Delta W=1$ across the 2nd-HL domain wall, producing a zero-frequency TDWS ($\omega^2/\omega_0^2=0$) labeled DW in Figs.~\ref{fig:interfacing_plot}(b,g) with their corresponding mode shape presented in Figs.~\ref{fig:interfacing_plot}(f,k). Notably, such 2nd-HL TDWSs can be realized by either (1) varying the inter-domain spacing $\frac{d}{D}$ while keeping $\frac{c_1}{c_2}$ fixed, or (2) inverting $\frac{c_1}{c_2}$ while preserving $\frac{d}{D}$. 

In contrast, Scenario $\it{iii}$ alters both $\frac{d}{D}$ and $\frac{c_1}{c_2}$ but leaves $\Delta W=0$. Therefore, no 2nd-HL TDWS forms. Instead, broken translational symmetry across the 2nd-HL domain wall introduces trivial defect states (D) inside the 1st-HL bandgap, appearing as asymmetric modes at $\omega^2/\omega_0^2\neq 0$ [Fig.~\ref{fig:interfacing_plot}(l,p)]. As this mode is not protected by the 2nd-HL topology, it never appears at $\omega^2/\omega_0^2 = 0$ at the center of the 2nd-HL bandgap.

The distinction between 1st-HL bulk modes (B1), 2nd-HL bulk modes (B2) within the 1st-HL bandgap, and 2nd-HL TDWS (DW) can be quantized using the participation ratio ($PR$) for each eigenmode $m$ in a finite system  \cite{thouless1974electrons,downing2024unconventional}, which is detailed in Section IV of the Supplementary Material.

Finally, the hierarchical construction can be extended: a third(3rd)-HL unit cell formed by pairing two 2nd-HL domain walls opens a new bandgap hosting a 3rd-HL TDWS at the interface between distinct 3rd-HL lattices. The number of such states is governed by a 3rd-HL winding number, $W$, which can be further generalized to any $n$th-HL (see Section II.A.1 of the main text for the generalized winding number expression and Section III of the Supplementary Material for detailed derivation and topological-state prediction application).

\subsubsection{Topological Domain Wall States Formed by Different Hierarchical Levels}

The ability to generate higher hierarchical topological states without dimension reduction enables TDWS formation across different hierarchical levels. Crucially, only a non-zero winding number difference ($\Delta W \neq 0$) is required, irrespective of the HLs involved, allowing for domain walls between, for example, 2nd-HL and 1st-HL lattices. 

\begin{figure*}
    \centering
    \includegraphics[scale=0.45]{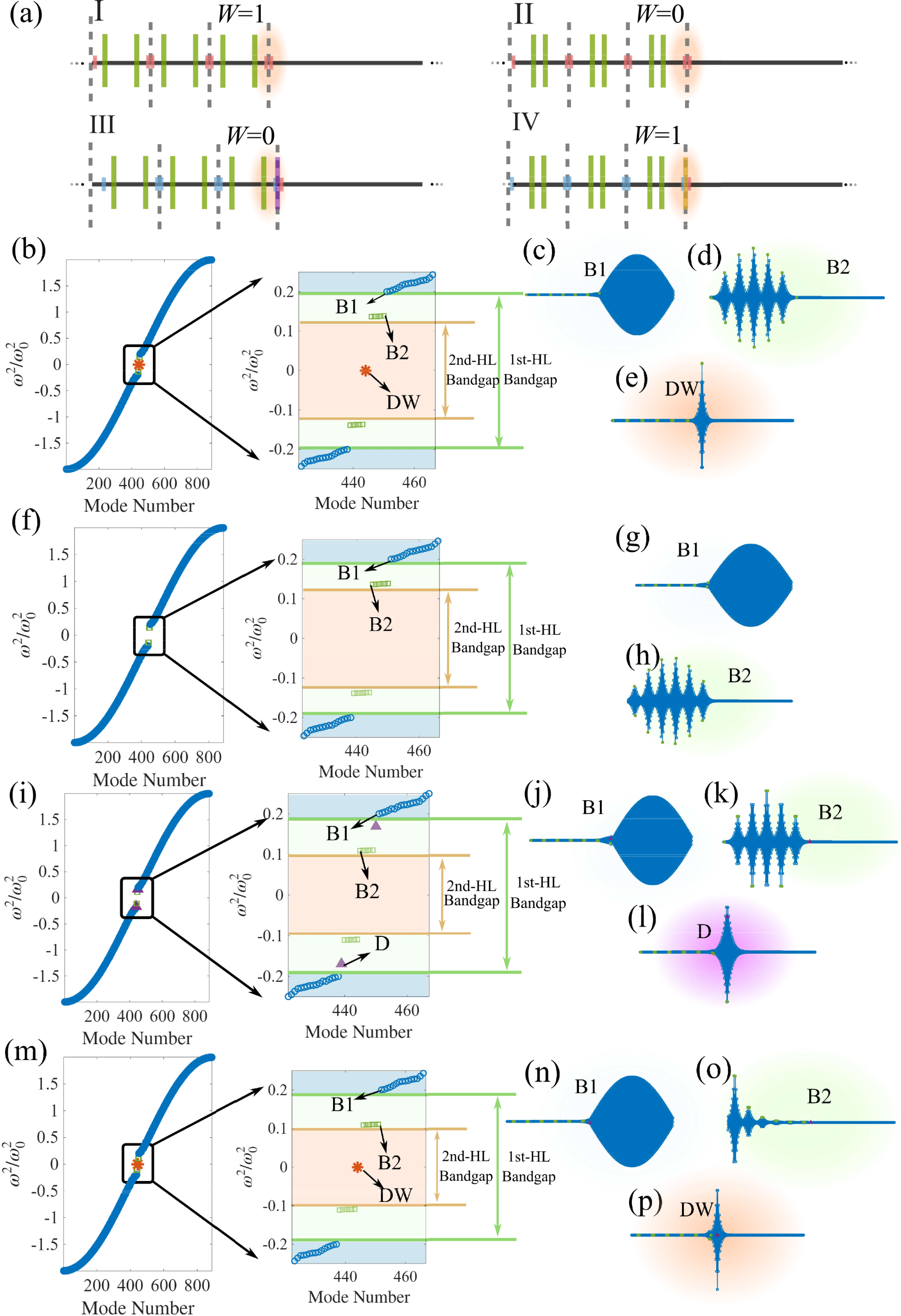} 
    \caption{Spectral and modal characteristics of one-dimensional mixed-hierarchical domain wall states. (a) Schematics of four finite lattices featuring mixed-HL domain walls (orange), each formed by six 2nd-HL unit cells (left) interfaced with 222 1st-HL unit cells ($c_1>c_2$). (b,f,i,m) Normalized eigenfrequencies ($\omega^2/\omega_0^2$) for the four configurations, with zoomed-in views (right panels) highlighting: 1st-HL bulk bands (B1, blue), 1st-HL bandgaps (green, hosting 2nd-HL bulk modes, B2), and 2nd-HL bandgaps (orange, potentially hosting 2nd-HL domain wall state, DW). (c-e,g-h,j-l,n-p) Representative mode shapes of B1, B2, topological DW states, and trivial defect modes (D). Green squares mark the 1st-HL domain walls.}
    \label{fig:mixed_order}
\end{figure*}

In our configuration, the right side consists of identical 1st-HL unit cells ($c_1>c_2$) with a zero 1st-HL winding number, while the left side alternates among the four 2nd-HL lattice types in Fig.~\ref{fig:W3D} (b) [schematics in] Fig.~\ref{fig:mixed_order} (a)]. Eigenvalue spectra [Fig.~\ref{fig:mixed_order} (b,f,i,m)], again, reveal consistent 1st- and 2nd-HL bandgaps across all four cases. 

Mixed-HL TDWSs emerge only when $\Delta W=1$ - specifically for Types I or IV 2nd-HL cells on the left - manifesting as zero-frequency modes ($\omega^2/\omega_0^2=0$) strongly localized at the mixed-HL interface [Fig.~\ref{fig:mixed_order} (e,p)], confirming their topological protection. In contrast, configurations with $\Delta W=0$ exhibit no such topological states [Fig.~\ref{fig:mixed_order} (f-k)]. 

As in the 2nd-HL domain wall case, broken translational symmetry in configuration III introduces trivial defect states (D) within the 1st-HL bandgap at $\omega^2/\omega_0^2\neq 0$ [Fig.~\ref{fig:interfacing_plot}(i,l)]. These modes, while still interface-localized, display slower spatial decay compared to their topological counterparts shown in Fig.~\ref{fig:mixed_order} (e,p), consistent with their proximity to the bulk band and non-topological origin.

\subsection{Two-Dimensional Hierarchical Su-Schrieffer-Heeger Lattices}
\subsubsection{Winding Number Calculation for Two-Dimensional Lattices}
\begin{figure*}
    \centering
    \includegraphics[scale=0.35]{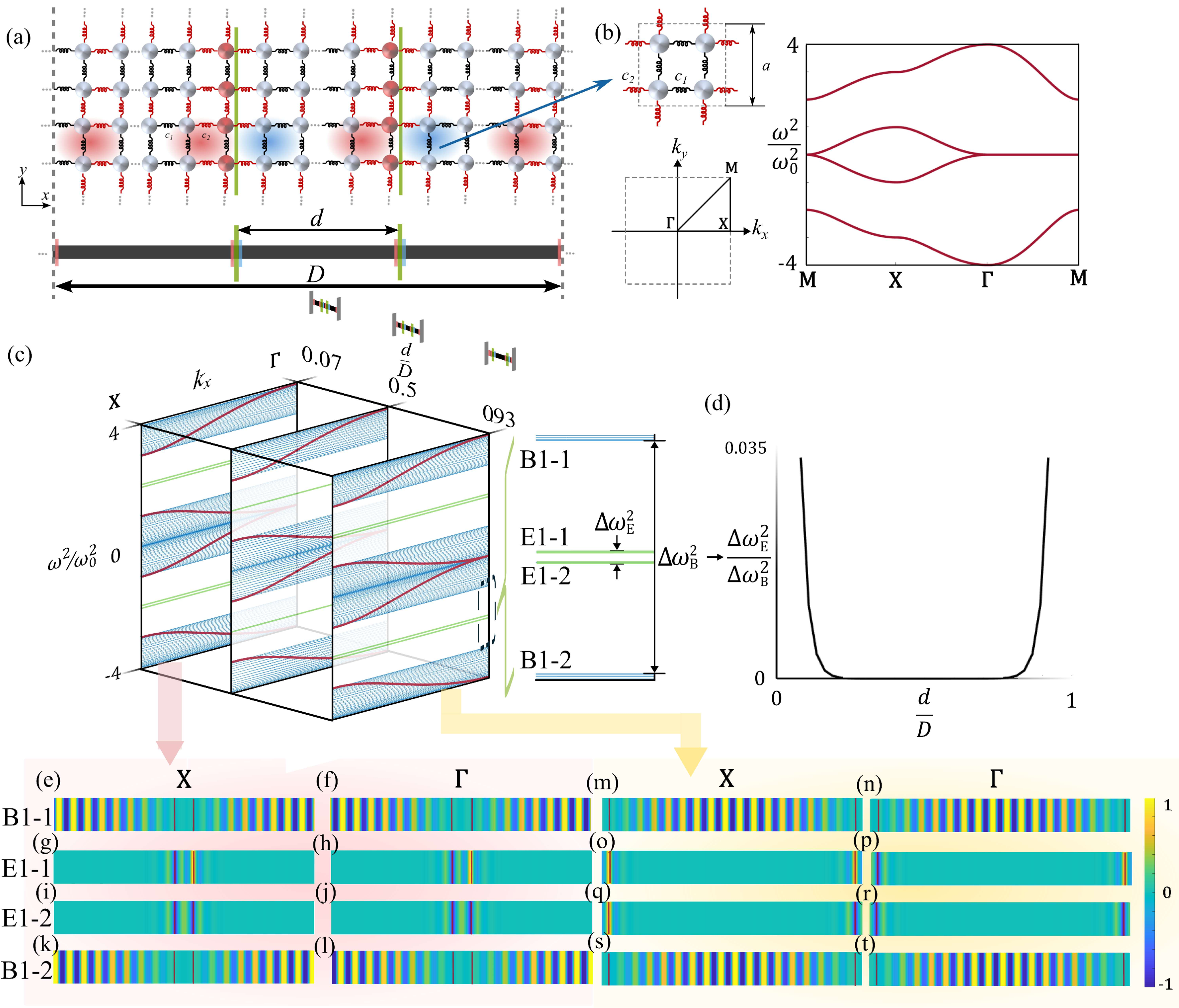} 
    \caption{Two-dimensional hierarchical lattice and its topological properties. (a) Schematic of a second-hierarchical-level (2nd-HL) lattice consisting of 74 masses in each row along the $x-$axis connected by springs with alternating stiffness ($c_1$, black; $c_2$, red). Red and blue shading highlights the simulated two rows of lattices with Bloch boundary conditions in both the $x$- and $y$-directions. Vertical green bars mark two first-hierarchical-level (1st-HL) domain walls. The 2nd-HL supercell, bounded by gray dashed lines, contains $d$ mass pairs between the walls and $D$ pairs in total, with a simplified representation below. (b) A 1st-HL unit cell ( top left), its reciprocal space with the irreducible Brillouin zone (bottom left), and its phonon dispersion (right). (c) Full phonon dispersion of the 2nd-HL system for $c_1=1.5$ and $c_2=0.5$ along $k_x$ (here, we denote it as X-$\mathrm\Gamma$ for easy comparison of modeshape parities) showing bulk bands (B1-1/2, blue), topological domain-wall states (TDWSs, E1-1/2 green), and 1st-HL bulk dispersions (red) for different $d/D$ ratios. (d) Normalized bandgap ratio $\Delta \omega_E^2/\Delta \omega_B^2$ between the 1st-HL TDWSs (E1-1/2) and bulk bands (B1-1/2) as a function of $d/D$. (e-t) Mode shapes at the $X$ and $\mathrm{\Gamma}$ points for bands B1-1/2 and E1-1/2, comparing configurations with (e-l) $d/D=0.07$ and (m-t) $d/D=0.93$. Red bars represent the two 1st-HL domain walls within each 2nd-HL cell.}
    \label{fig:2D_Interface_distance}
\end{figure*}

The hierarchical framework developed for 1D SSH lattices extends naturally to higher dimensions. In this section, we illustrate this using a 2D SSH model as an example. As shown in Fig.~\ref{fig:2D_Interface_distance} (a,b), the lattice consists of identical masses connected by alternating springs $c_1$ and $c_2$ along both $x-$ and $y-$directions. The effective Hamiltonian of a 1st-HL unit cell $\bm{H}^\prime(\mathbfit{k})$, equivalent to a tight-binding model and derived from the stiffness matrix, $\bm{H}(\mathbfit{k})$ (detailed in Section V of the Supplementary Material), becomes:
\begin{equation}
	\bm{H}^\prime(\mathbfit{k}) = 
	\begin{bmatrix}
		0 & h_{12}^\dagger & h_{13}^\dagger & 0 \\
		h_{12} & 0 &0 & h_{24}^\dagger \\
        h_{13} &0 &0 & h_{34}^\dagger \\
        0 & h_{24} & h_{34} & 0
	\end{bmatrix}
	\label{eq:2D_Hmatrix}
\end{equation}
where $h_{12}=-c_1-c_2e^{ik_xa}$, $h_{13}=-c_1-c_2e^{ik_ya}$, $h_{24}=-c_1-c_2e^{ik_ya}$, $h_{34}=-c_1-c_2e^{ik_xa}$, and $\mathbfit{k}=(k_x,k_y)$. Unlike the 1D SSH model, the 2D variant exhibits an indirect bandgap at $\omega^2/\omega_0^2\neq0$. Additionally, the symmetry of the system enforces $h_{12}=h_{34}$ and $h_{13}=h_{24}$, reflecting equivalent couplings along $x-$ and $y-$directions.

Although the topology of this 2D SSH model can be characterized by a 2D polarization - analogous to the Zak phase in 1D and computed from the eigenvectors of the basic (1st-HL) unit cell~\cite{liu2017novel,xie2018second} - this approach becomes increasingly challenging for higher-HL lattices due to Hamiltonian complexity and frequent band crossings that obscure topological invariants. Instead, we compute winding numbers along each axis using Eqn.~\ref{winding_number_original}. This is justified by the separable form of the Hamiltonian in Eqn.~\ref{eq:2D_Hmatrix}, which decomposes into independent dynamics along the $x-$ and $y-$axes: 
\begin{equation}
\bm{H}^\prime(\mathbfit{k})=\bm{I}\otimes \bm{H}_x^\prime+\bm{H}_y^\prime\otimes \bm{I},
\label{seperableH}
\end{equation}
where $\bm{H}_x^\prime= \begin{pmatrix} 0 & h_{12}^\dagger \\ h_{12} & 0 \end{pmatrix}$ and $\bm{H}_y^\prime= \begin{pmatrix} 0 & h_{13}^\dagger \\ h_{13} & 0 \end{pmatrix}$. Substituting $\bm{H}^\prime$ with $\bm{H}_x^\prime$ or $\bm{H}_y^\prime$ in Eqn.~\ref{winding_number_original} yields $W=1$ when $c_1<c_2$ and $W=0$ when $c_1>c_2$ in both directions, indicating the potential for TDWSs along $x$ and $y$. Given the symmetry between directions, we focus on hierarchical lattice construction along the $x$-axis, while maintaining consistent $c_1$ and $c_2$ arrangements along the $y$-direction. Therefore, two rows of lattices with Bloch conditions applied along $y$ to represent lattices infinitely long along the $y$-direction.

Analogous to the construction of 2nd-HL cells in 1D SSH models, we initiate the process by forming one TDWS through a supercell comprising a single domain wall, where the lattices in the two domains exhibit opposite $c_1$ and $c_2$ arrangements. This configuration hosts a flat TDWS at the bandgap center along the X-$\mathrm{\Gamma}$ path in the $x$-direction ($\omega^2/\omega_0^2=\pm2$),  corresponding to $\Delta W=1$ between the two lattice phases about the domain wall. Doubling the supercell introduces two domain walls, producing two degenerate TDWSs. Tuning their separation ($\frac{d}{D}\neq\frac{1}{2}$) splits these states [Fig.~\ref{fig:2D_Interface_distance} (c,d)]. Modeshape analysis at X and $\mathrm{\Gamma}$ [Fig.~\ref{fig:2D_Interface_distance} (e-t)] reveals that for $\frac{d}{D}<\frac{1}{2}$, no parity inversion occurs, indicating a topologically trivial bandgap ($W=0$) between the two TDWSs. Conversely, when $\frac{d}{D}>\frac{1}{2}$, both TDWSs exhibit parity inversion, signifying a topological bandgap ($W=1$). Since these TDWSs are flat along $k_x$, these states are localized at the domain walls without propagating along $x$. 

In contrast, dispersion along $k_y$ reveals two in-gap propagating states (with non-zero group velocities). At $\mathrm{\Gamma}$, the TDWSs are located near the center of the bandgap ($\omega^2/\omega_0^2=\pm2$). Adjusting  $\frac{d}{D}$ away from $\frac{1}{2}$ opens a gap between these states [Fig.~\ref{fig:2dunit_supp} (a-d)]. However, the modal analysis suggests no parity inversion occurs at X and $\mathrm{\Gamma}$ either when $\frac{d}{D}<\frac{1}{2}$ or $\frac{d}{D}>\frac{1}{2}$. This further confirms that the hierarchical topology along $x$ does not affect the $y$-direction. This observation is consistent with the aforementioned separability of $\bm{H}^\prime$ in Eqn.~\ref{eq:2D_Hmatrix}, implying that hierarchical winding numbers along each direction can be computed independently from the reduced Hamiltonians, $\bm{H}^\prime_x$ and $\bm{H}^\prime_y$. 

Following the procedure established for the 1D case in Sections II of the Supplementary Material, which leads to the final expression in Eqn.~\ref{eq:gen_W_express}, we confirm that for configurations with $c_1>c_2$, the winding number satisfies $W=0$ for $\frac{d}{D}<\frac{1}{2}$ and $W=1$ for $\frac{d}{D}>\frac{1}{2}$. Higher-level hierarchical winding numbers can also be derived analogously as described in Section III of the Supplementary Material, yielding the generalized $W_n$ and $\abs{A}_n$ in Eqns.~\ref{eq:Wn} and~\ref{eq:An} in each axis.

The following subsections examine the impact of 2nd-HL winding numbers on edge states, TDWSs, and domain walls connecting mixed hierarchical levels in 2D hierarchical lattices. All configurations considered here consist of two rows of 2nd-HL unit cells coupled by springs $c_1$ with Bloch boundary conditions (via $c_2$) imposed along the $y-$axis to emulate an infinitely long lattice with alternating $c_1$ and $c_2$ couplings. All the dispersion curves are plotted along the $k_y$ direction. With each row, the spring arrangements follow the notation in Fig.~\ref{fig:W3D}. Each lattice contains $D=37$ mass pairs in each row, with the number of 1st-HL unit cells between domain walls (green bars) given by $d=34$ (Type I), $d=2$ (Type II), $d=33$ (Type III), and $d=3$ (Type IV).

\subsubsection{Second-Hierarchical Topological Edge States of a Finite
Lattice}
Analogous to the 1D cases discussed in Section II B.2, we construct four finite lattices composed of 2nd-HL cells, each mass row corresponding to the categories in Fig.~\ref{fig:W3D} with appropriate boundary termination. To preserve chirality, additional springs are introduced at domain walls and terminations so that every mass retains an effective stiffness of $2(c_1 + c_2)$. 
\begin{figure*}
	\centering
	\includegraphics[scale=0.22]{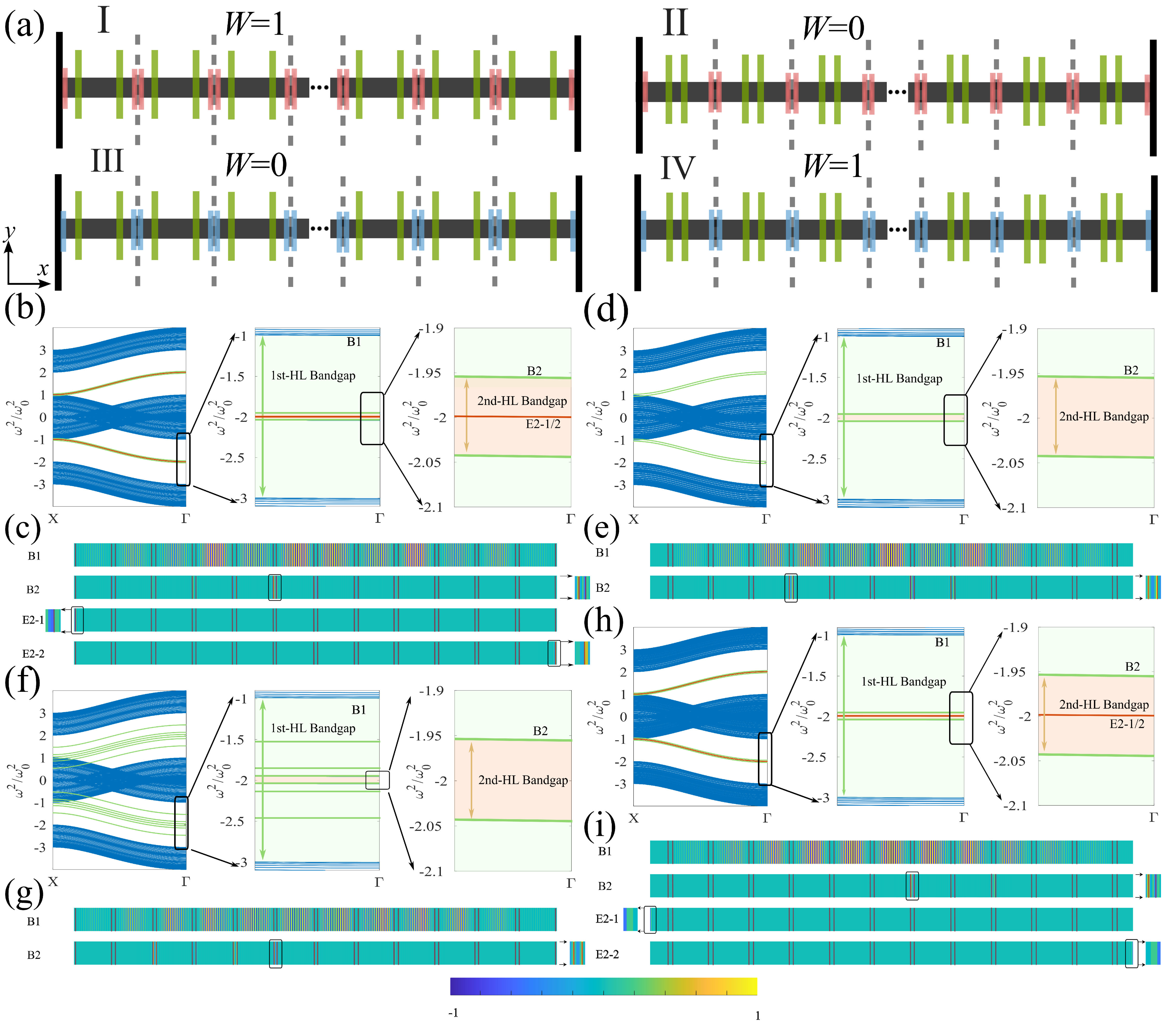} 
	\caption{Spectral and modal characteristics of two-dimensional finite hierarchical lattices. (a) Schematics of four finite lattices, each comprising two connected rows of 12 second-hierarchical-level (2nd-HL) unit cells. Bloch boundary conditions are applied along the $y$-axis to emulate an infinitely long lattice connected by alternating springs $c_1$ and $c_2$ in that direction. Black bars indicate boundary masses with additional grounding springs to maintain a total stiffness of $2(c_1+c_2)$ for all masses. (b,d,f,h) Normalized eigenfrequencies ($\omega^2/\omega_0^2$) for lattices I-IV in (a) along the $k_y$ direction [$i.e.$, the X-$\mathrm{\Gamma}$ path in the unit cell shown in Fig.~\ref{fig:2D_Interface_distance}(b)]. Middle and right panels provide zoomed-in views of key spectral features: 1st-HL bulk modes (B1, blue), 1st-HL bandgaps (green, containing 2nd-HL bulk modes, B2), and 2nd-HL bandgaps (orange, potentially hosting 2nd-HL edge states, E2-1/2). (c,e,g,i) Representative mode shapes for B1, B2, and E2-1/2 (if present) modes for each lattice. Red vertical bars mark the 1st-HL domain wall masses.}
	\label{fig:fixed_boundary_2d}
\end{figure*} 

The resulting lattices are shown in Fig.~\ref{fig:fixed_boundary_2d} (a), each based on a distinct 2nd-HL cell type [Fig.~\ref{fig:W3D} (b)]. Eigenvalue spectra along $k_y$ exhibit consistent 1st- and 2nd-HL bulk bandgaps across all cases [Fig.~\ref{fig:fixed_boundary_2d} (b,d,f,h)]. As in the 1D cases (Fig.~\ref{fig:fixed_boundary}), TPESs at the termination boundaries at $\omega^2/\omega_0^2=\pm2$ appear only for lattices containing Types I and IV cells [Fig.~\ref{fig:fixed_boundary_2d} (c,i)], and are absent for Types II and III [Fig.~\ref{fig:fixed_boundary_2d} (e,g)], consistent with their repsecitve winding numbers.

Unlike in 1D, where 2nd-HL bulk and edge modes are both static and localized, the corresponding modes in 2D propagate along the $y$-axis, resulting in a nonzero group velocity $v_g$ along the X-$\mathrm{\Gamma}$ path.

\subsubsection{Second-Hierarchical Topological Domain-Wall States}

\begin{figure*}
	\centering
	\includegraphics[scale=0.22]{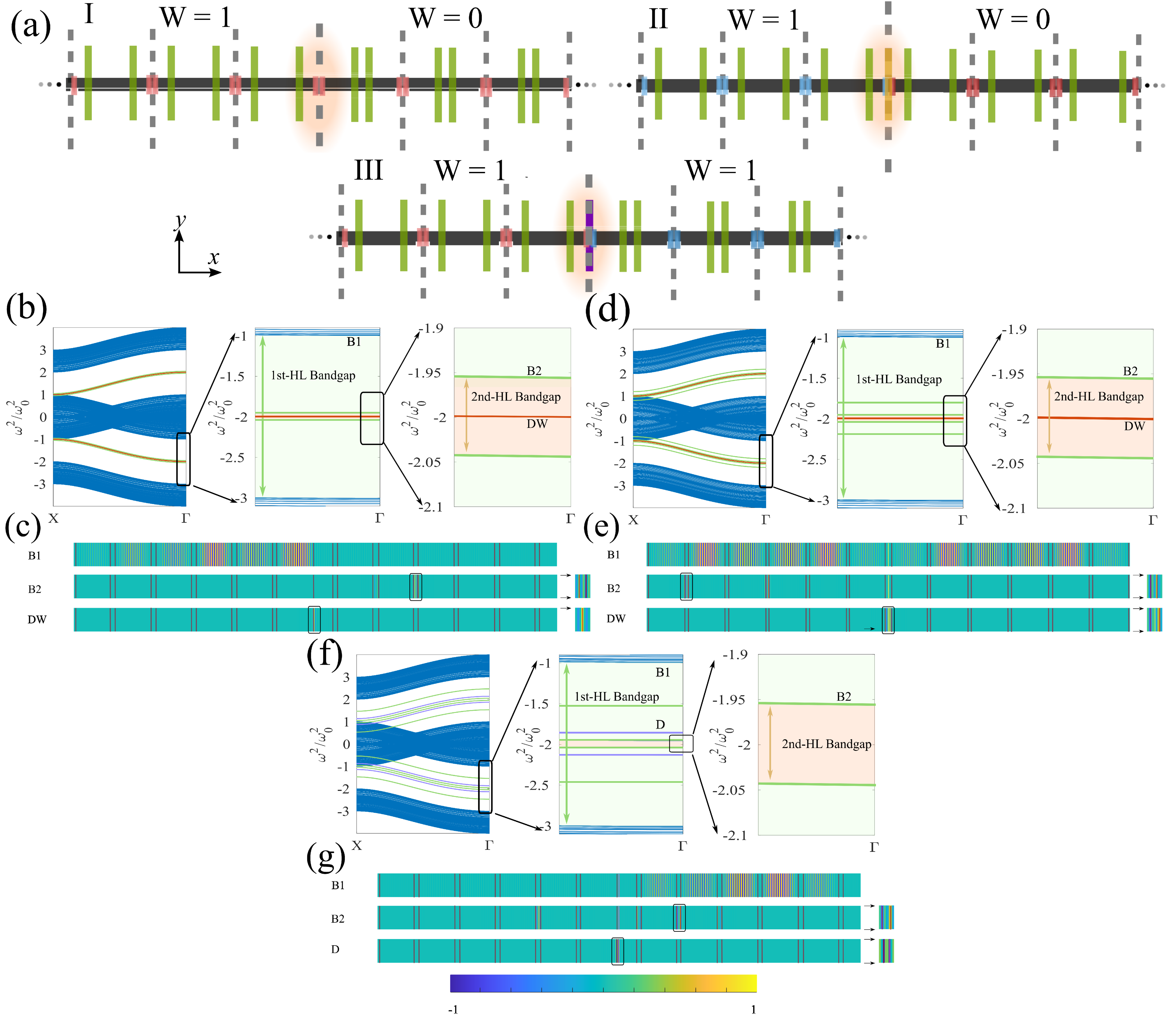} 
	\caption{Spectral and modal characteristics of two-dimensional hierarchical domain wall states. (a) Schematics of three second-hierarchical-level (2nd-HL) domain wall configurations, each comprising two rows of 6 2nd-HL cells per side, with terminations along $x$ and Bloch boundary conditions along $y$ to emulate an infinitely long lattice connected by alternating springs $c_1$ and $c_2$ in that direction. (b,d,f) Normalized eigenfrequencies ($\omega^2/\omega_0^2$) along $k_y$ (X-$\mathrm{\Gamma}$) for Scenarios $\it{i}$-$\it{iii}$ in (a), with zoomed-in views highlighting 1st-HL bulk modes (B1, blue), 1st-HL bandgaps (green, containing 2nd-HL bulk modes, B2), and 2nd-HL bandgaps (orange, potentially hosting 2nd-HL domain wall state, DW), and trivial defect modes within 1st-HL bandgaps (purple, D). (c,e,g) Representative mode shapes of B1 B2-1/2, DW, and D for each scenario. Red and yellow/purple bars indicate 1st- and 2nd-HL domain wall masses, respectively.}
	\label{fig:interfacing_plot_2D}
\end{figure*}

We consider three distinct domain wall configurations, analogous to the 1D cases in Fig.~\ref{fig:interfacing_plot} (a), to verify the number of 2nd-HL TDWSs predicted by the winding numbers (Eqn.~\ref{eq:gen_W}) in 2D lattices. These domain walls join: $i$ Types I ($W=1$) and II ($W=0$), $ii$ III ($W=0$) and I ($W=1$), and $iii$ I ($W=1$) and IV ($W=1$). The corresponding structures, eigenfrequencies, and eigenmodes along $k_y$ (X-$\mathrm{\Gamma}$) are shown in Fig.~\ref{fig:interfacing_plot_2D}. In all cases, 2nd-HL bandgaps emerge because $\frac{d}{D}\neq \frac{1}{2}$ on both sides of the 2nd-HL domain wall, bounded by 2nd-HL bulk modes that effectively represent the simultaneous excitation of multiple 1st-HL TDWSs [Fig.~\ref{fig:interfacing_plot} (c,e,g)]. These higher-HL bandgaps enable the formation of higher-HL topological states.

In Scenarios $\it{i}$ and $\it{ii}$, the combined effects of $\frac{d}{D}$ and $\frac{c_1}{c_2}$ yield a winding number difference $\Delta W=1$, producing a TDWS at $\omega^2/\omega_0^2=\pm 2$ [Figs.~\ref{fig:interfacing_plot_2D}(b-e)]. Notably, such 2nd-HL TDWSs can be induced either by (1) varying $\frac{d}{D}$ across $\frac{1}{2}$ while keeping $\frac{c_1}{c_2}$ fixed or (2) inverting the $c_1$ and $c_2$ arrangement while preserving $\frac{d}{D}$. 

By contrast, Scenario $\it{iii}$ shows that when both $\frac{d}{D}$ and $\frac{c_1}{c_2}$ change simultaneously, the winding numbers remain equal on both sides of the domain wall ($\Delta W=0$), and no 2nd-HL TDWSs form, despite the 2nd-HL unit cells having different configurations. Instead, broken translational symmetry introduces trivial defect states (D) within the 1st-HL bandgap, appearing as asymmetric modes at $\omega^2/\omega_0^2\neq\pm 2$ [Fig.~\ref{fig:interfacing_plot}(f,g)]. As these states lack 2nd-HL topological protection, they never occur at the center of the 2nd-HL bandgap, $i.e.,$ $\omega^2/\omega_0^2 = \pm 2$.

\subsubsection{Topological Domain Wall States Formed by Different Hierarchical Levels}

Analogous to the 1D case, the ability to construct higher hierarchical topological states without dimension reduction enables the realization of TDWSs through two distinct hierarchical levels. Importantly, TDWS formation only requires a non-zero winding number difference, $i.e.$, $\Delta W \neq 0$, irrespective of the specific HLs involved. This permits domain walls to combine, for example, by interfacing a 2nd-HL lattice on one side with a 1st-HL one on the other. 

\begin{figure*}
    \centering
    \includegraphics[scale=0.22]{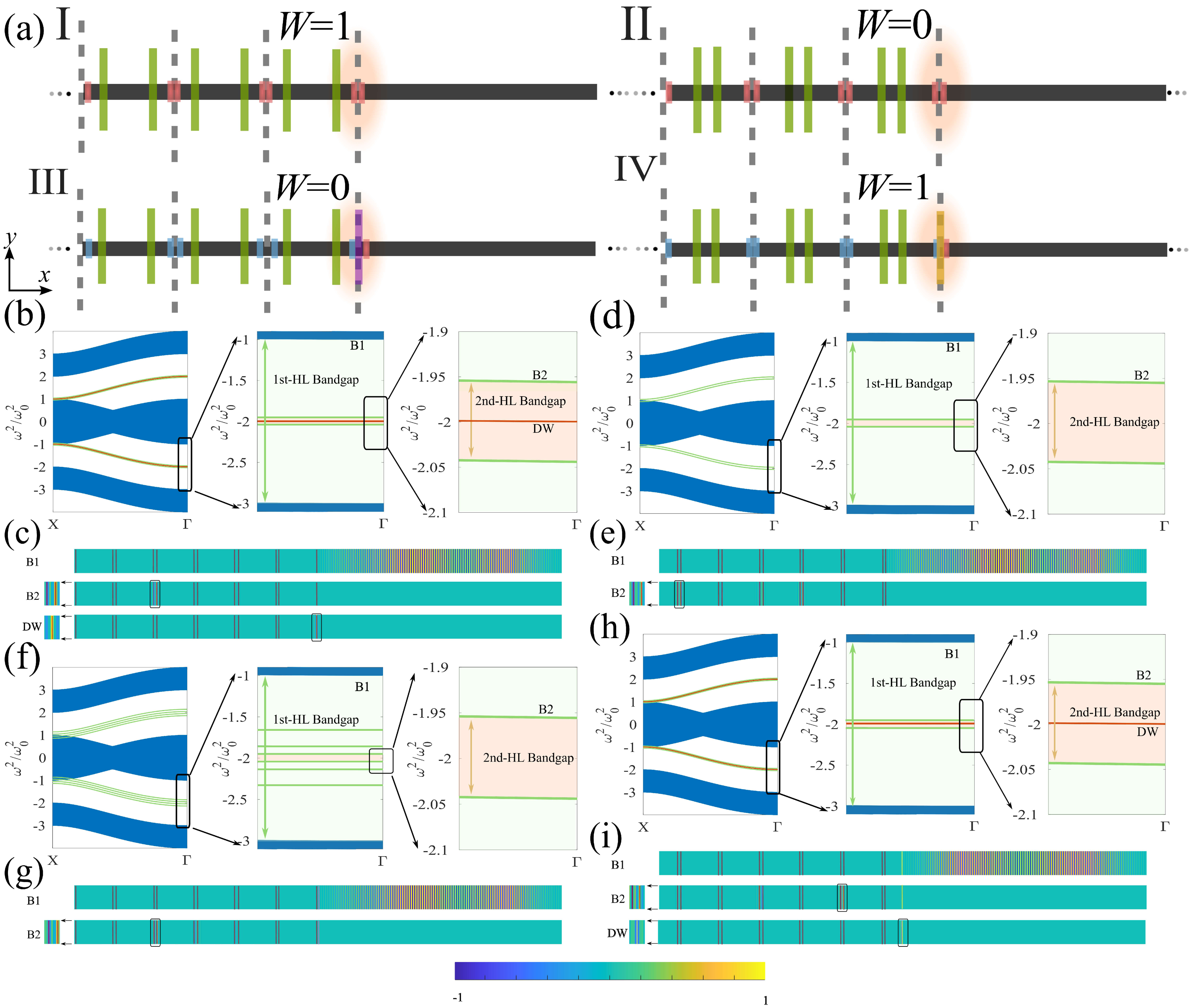} 
    \caption{Spectral and modal characteristics of two-dimensional mixed-hierarchical domain wall states. (a) Schematics of four mixed-HL configurations, each formed by two rows of six second-HL cells (left) interfaced with 222 first-HL cells (right, all with $c_1>c_2$), terminated along $x$ and periodic along $y$ to emulate an infinitely long lattice connected by alternating springs $c_1$ and $c_2$ in that direction. (b,f,i,m) Normalized eigenfrequencies ($\omega^2/\omega_0^2$) along $k_y$ (X-$\mathrm{\Gamma}$), with insets highlighting 1st-HL bulk bands (B1, blue), 1st-HL bandgaps (green, containing 2nd-HL bulk modes, B2), and 2nd-HL bandgaps (orange, potentially hosting domain wall state, DW). (c-e,g-h,j-l,n-p) Representative mode shapes of 1st-HL bulk (B1), 2nd-HL bulk (B2), topological domain-wall states (DW), and trivial defect modes (D). Red and yellow/purple bars indicate 1st- and 2nd-HL domain wall masses, respectively.}
    \label{fig:mixed_order_2D}
\end{figure*}

In our design, the right side of the mixed-HL domain wall consists of identical 2D 2nd-HL unit cells with $c_1>c_2$ in each row, ensuring a zero 1st-HL winding number in the $x$-direction. On the left, we systematically vary the 2nd-HL lattices among the four lattice types shown in Fig.~\ref{fig:W3D} (b), extended to 2D with Bloch boundary conditions applied in the $y$-direction [Fig.~\ref{fig:mixed_order_2D} (a)]. Eigenvalue spectra along $k_y$ ($i.e.$, X-$\mathrm{\Gamma}$), reveal robust 1st- and 2nd-HL bandgaps across all configurations [Fig.~\ref{fig:mixed_order_2D} (b,f,i,m)]. It turns out that mixed-HL TDWSs that propagate along the $y$-direction appear only when $\Delta W=1$, specifically, for Types I or IV on the left, as shown in Fig.~\ref{fig:mixed_order_2D} (e,p). These states are exactly at $\omega^2/\omega_0^2=\pm 2$ and exhibit strong spatial localization at the mixed-HL domain walls, a hallmark of topological protection. In contrast, configurations with $\Delta W=0$ host no such topological states at the mixed-HL domain walls [Fig.~\ref{fig:mixed_order_2D} (f-k)]. 

As in the 2nd-HL domain wall case, broken translational symmetry in configuration III [Fig.~\ref{fig:mixed_order_2D} (a)] introduces trivial defect modes (D) within the first-HL bandgaps at $\omega^2/\omega_0^2\neq \pm 2$ [Fig.~\ref{fig:interfacing_plot_2D}(i,l)]. While still localized at the mixed-HL interface, these defect modes decay more slowly in space than their topological counterparts [Fig.~\ref{fig:mixed_order_2D} (e,p)], consistent with their proximity to the bulk band. 

\section{\label{sec:level3} Conclusion}

In summary, we introduce a general framework for generating hierarchical topological states in 1D and 2D Su-Schrieffer-Heeger models without breaking the symmetry of the original topological states ($i.e.$, further breaking the basic unit cell symmetry) or resorting to dimension reduction. Higher-hierarchical-level topology emerges from bandgap openings between degenerate lower-hierarchical-level topological states—a process controlled by strategically positioning domain walls. This new mechanism extends topological hierarchies beyond conventional classifications, which requires further symmetry breaking of the topological states and a reduction in dimensions. Moreover, the topology characterization is supported by a generalized winding number formalism applicable to multiple-hierarchical-level systems, overcoming the constraints of diatomic models in conventional winding number calculation.

Our analysis confirms the existence of 2nd-HL topological domain-wall and edge states and demonstrates the preservation of bulk-edge correspondence across hierarchies. The approach is inherently scalable to higher dimensions, offering a versatile platform for hierarchical metamaterials. The robustness of these states against large perturbations makes them promising for applications requiring precise vibration localization, such as phonon-based quantum information~\cite{ma2023phonon} processing and engineered structures for vibration control~\cite{zhang2013broadband,mousanezhad2015honeycomb}. These results establish a foundation for programmable topological responses in complex systems.

\section{Acknowledgment}
J.R.P. acknowledges the support of the National Aeronautics and Space Administration (NASA Grant 80NSSC20M0122). J.A.M. acknowledges the support of the National Aeronautics and Space Administration (NASA Grant 80NSSC23M0071).

\bibliography{bib}
\pagebreak
\widetext
\begin{center}
	\textbf{\large Supplemental Material: Dimensionless Hierarchical Topological Phononic States}
\end{center}
\setcounter{equation}{0}
\setcounter{figure}{0}
\setcounter{table}{0}
\setcounter{page}{1}
\setcounter{section}{0}
\makeatletter
\renewcommand{\theequation}{S\arabic{equation}}
\renewcommand{\thefigure}{S\arabic{figure}}

\section{One-Dimensional Su-Schrieffer-Heeger Model}
\begin{figure*}[!htbp]
	\centering
	\includegraphics[scale=0.55]{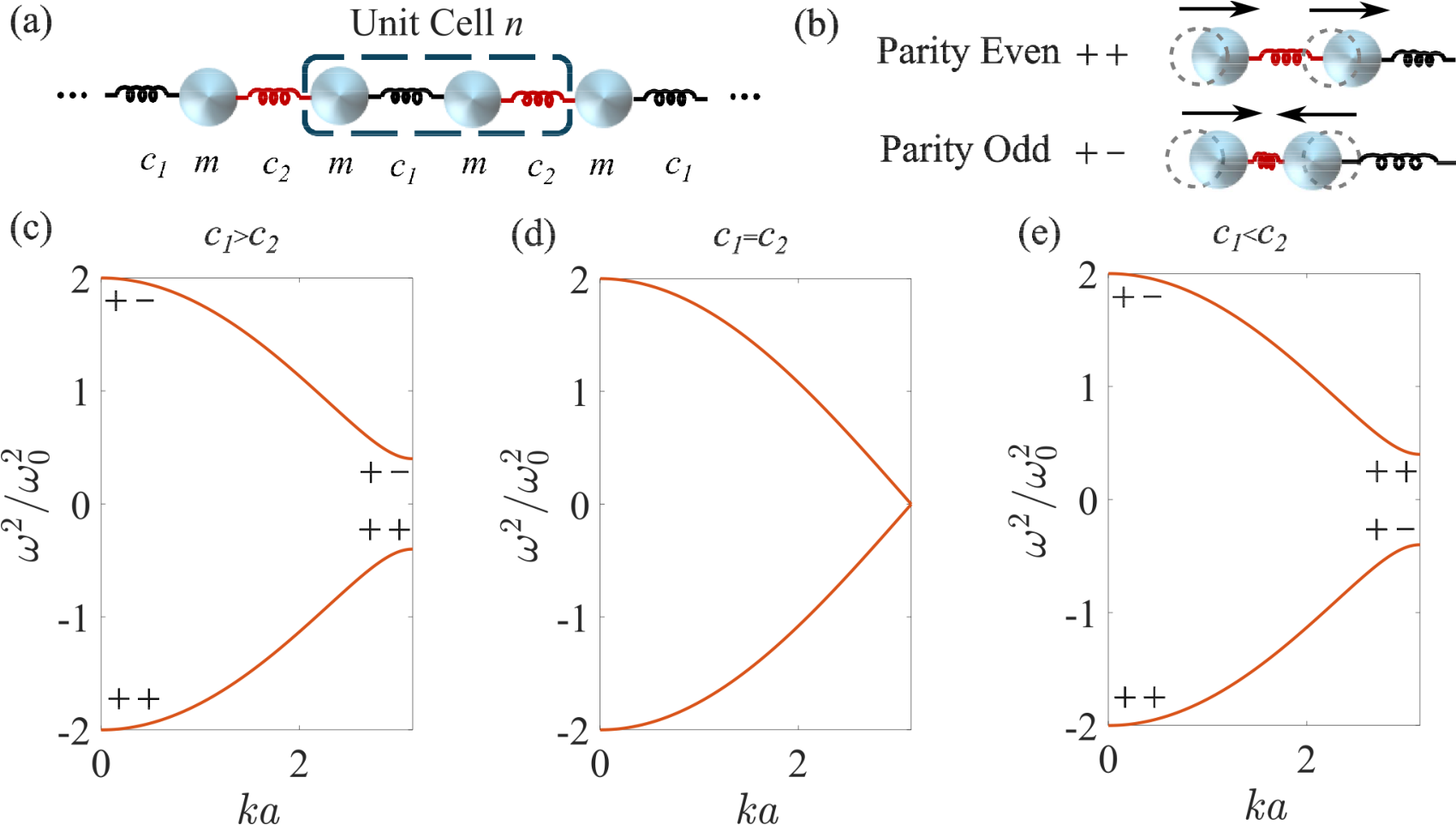} 
	\caption{(a) Unit cell (encircled by a dashed line) of a Su-Schrieffer-Heeger chain consisting of identical masses, $m$, and alternating springs with spring constants, $c_1$ and $c_2$. (b) Possible parities of unit cell mode shapes with both masses moving in the same direction ($i.e.$, parity even), labeled as "++", and in the opposite directions ($i.e.$, parity odd), labeled as "+-." Dashed circles denote the mass positions without displacements. (c-e) Phonon dispersion relations of the unit cell with (c) $c_1>c_2$, (b) $c_1=c_2$, and (c) $c_1<c_2$, with parities at $k=0$ and $k=\pi/a$ denoted in the plots, where $k$ is the wave number and $a$ is the lattice vector. }
	\label{fig:S1}
\end{figure*}

The one-dimensional (1D) mechanical Su-Schrieffer-Heeger (SSH) model examined in this work is constructed using identical masses and springs of alternating strength, as illustrated in Fig.~\ref{fig:S1}(a). The equations of motion for this system can be derived using Newton's Second Law:

\begin{eqnarray}
	m\ddot{u}_1^n = c_1(u_2^n - u_1^n) - c_2(u_1^n - u_2^{n-1}), \nonumber \\
	m\ddot{u}_2^n = c_2(u_1^{n+1} - u_2^n) - c_1(u_2^n - u_1^n),
	\label{eq:spring_mass_gov}
\end{eqnarray}
where $u_1^n$ and $u_2^n$ represent displacements of the two masses in the $n^{th}$ unit cell, and can be combined using a plane-wave solution in combination with Bloch periodic boundary conditions:

\begin{eqnarray}
	{\mathbfit u}^n(t)={\tilde{\mathbfit{u}}}(k)e^{i(nka-\omega t)},
	\label{eq:planewave}
\end{eqnarray}
where $\omega$ is the vibration frequency, ${\mathbfit u}^n$ are the displacements of the $n$-th cell with $\mathbfit{u}^n=[u_1^n,u_2^n]^\mathrm{T}$, $k$ is the wave number, which is inversely proportional to the wavelength $\lambda$, \textit{i.e.}, $k=2\pi/\lambda$, $a$ denotes the lattice constant, $\tilde{\mathbfit{u}}(k)$ are displacements within the unit cell. Substituting this expression in Eqns.~\ref{eq:spring_mass_gov} gives:
\begin{eqnarray}
	[\bm{H}(k)-\omega^2m]{\tilde{\mathbfit{u}}}(k)=0,
	\label{goveqn}
\end{eqnarray}
where $\bm{H}(k)$ is the stiffness matrix of the periodic system:
\begin{equation}
	\bm{H}(k) = 
	\begin{bmatrix}
		c_1+c_2 & -c_1 - c_2e^{-ika} \\
		-c_1 - c_2e^{ika} & c_1+c_2
	\end{bmatrix}
	\label{eq:Hmatrix}
\end{equation}
Since the diagonal elements are identical and are responsible for shifting the eigenvalues, we can remove them to obtain a chiral matrix resembling a Hamiltonian of a generalized tight-binding model:
\begin{eqnarray}
	\bm{H}'(k)=\bm{H}(k)-(c_1+c_2)\bm{I},
	\label{eq:chiral}
\end{eqnarray}
where $\bm{I}$ is the identity matrix. In the tight-binding model, $c_1$ and $c_2$ are the intra- and inter-cell hopping parameters, respectively. By solving for the eigenvalues of $\bm{H}'(k)$, one can derive a band structure that is symmetric about the zero-frequency (or zero-energy, in quantum systems) state, $i.e.$, $\frac{\omega^2}{\omega_0^2} = 0$, where \( \omega_0^2 = \frac{c_1 + c_2}{m} \), as illustrated in Fig. \ref{fig:S1}(c-e). It is important to note that we can achieve the same bandgap by swapping the values of \( c_1 \) and \( c_2 \). However, when \( c_1 < c_2 \), the parities of the unit-cell eigenvectors are exchanged at \( k = \frac{\pi}{a} \), which indicates a topological bandgap. This is in contrast to the situation when \( c_1 > c_2 \).

\begin{figure}[!htbp]
	\centering
	\includegraphics[scale=0.52]{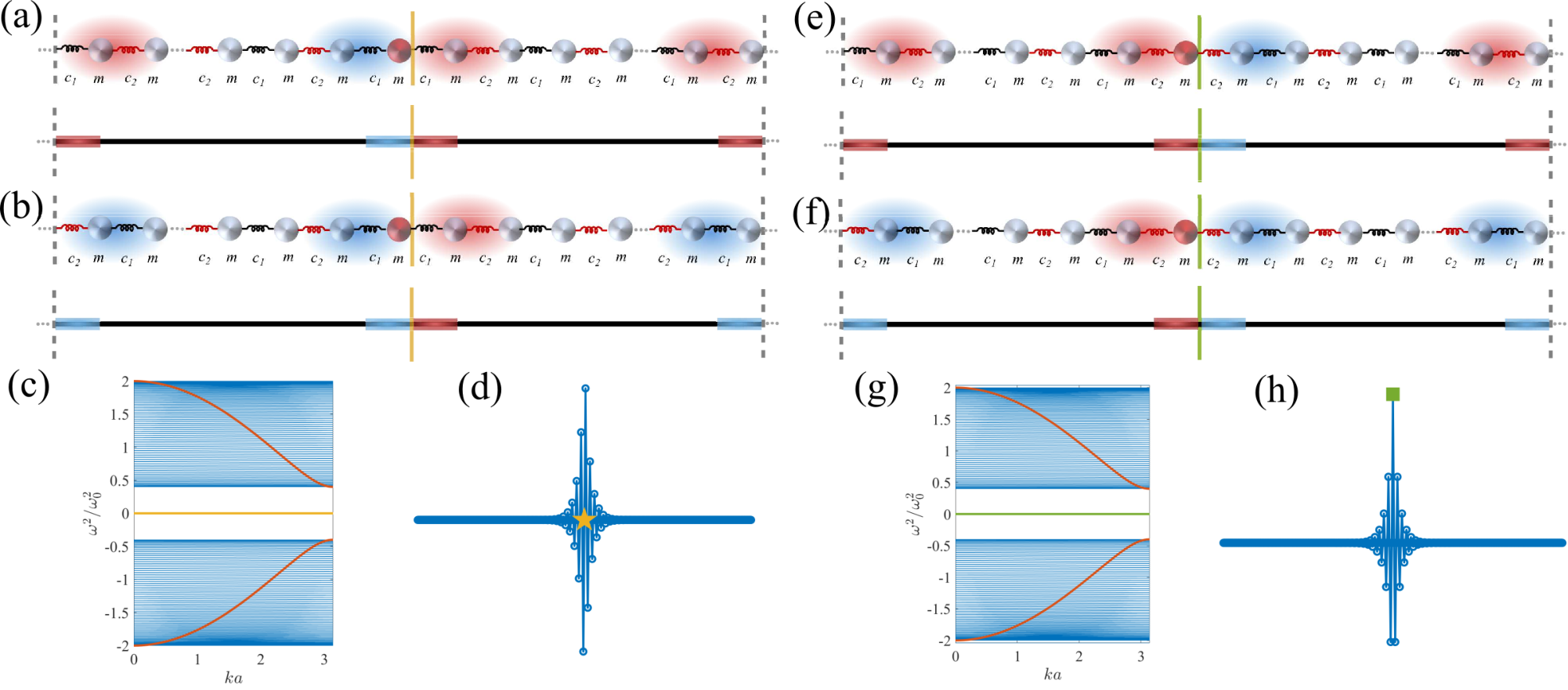} 
	\caption{(a,b,e,f) Supercells with periodic boundary conditions extending beyond the dashed truncation lines at both ends featuring two types of interfaces with the interfacial mass (highlighted in red) connected to (a,b) $c_1$ and (e,f) $c_2$, respectively. (a,e) and (b,f) display two types of terminating cells, indicated by red and blue shading, respectively. Blue curves in (c,g) represent the phonon dispersion relations for the supercells shown in (a,b) and (e,f), respectively, with the topological domain wall states (TDWSs) highlighted in yellow and green in each plot. The red curves indicate the phonon dispersions of the superimposed unit cells. (d) and (h) present the mode shapes of the TDWSs with the yellow star and green square marking the locations of the interfacial masses in (c) and (g), respectively. The vertical displacements in the modeshapes represent the horizontal displacements for clearer visualization. In all cases, $c_1>c_2$.}
	\label{fig:S2}
\end{figure}

Joining these two phases creates a topological domain-wall state (TDWS) at \(\omega^2/\omega_0^2=0\), also referred to as the zero-frequency state, as illustrated in Fig.~\ref{fig:S2}. The mode shape symmetries of the TDWS can vary depending on the arrangement of springs around the domain wall. However, this variation is not influenced by the truncating topology of the unit cells within the supercell. 

\section{Topology of the One-Dimensional Second-Hierarchical-Level Unit Cell}
As we expand the unit cell beyond the diatomic system containing $N$ pairs of diatomic unit cells, the stiffness matrix, $\bm{H}$ becomes:

\begin{equation}
	\bm{H} = 
	\begin{bmatrix}
		c_1 + c_2 & -c_1 & 0 && 0&0& -c_2e^{-ika}  \\
		-c_1 & c_1+c_2 & -c_2  && 0&0&0 \\
		 &\vdots&&\ddots&&\vdots&\\
		0 &0& 0&&-c_2 & c_1 + c_2  & -c_1\\
		-c_2e^{ika} & 0&0&&0&-c_1 & c_1 + c_2 \\
	\end{bmatrix}_{2N\times 2N}\label{S_Hmatrix}
\end{equation}  
By subtracting the identical diagonal elements, as shown in Eqn.~\ref{eq:chiral}, we can get the chiral matrix resembling a Hamiltonian of a tight-binding model, $\bm{H}'$:

\begin{equation}
	\bm{H}' = 
	\begin{bmatrix}
	0 & -c_1 & 0 && 0&0& -c_2e^{-ika}  \\
	-c_1 & 0 & -c_2  && 0&0&0 \\
	&\vdots&&\ddots&&\vdots&\\
	0 &0& 0&&-c_2 & 0  & -c_1\\
	-c_2e^{ika} & 0&0&&0&-c_1 & 0 \\
\end{bmatrix}_{2N\times 2N}\label{S_H'}
\end{equation} 

To obtain the general winding number presented in Eqn.~\ref{eq:gen_W}, the third Pauli matrix, $\bm{\sigma}_3$, in Eqn.~\ref{winding_number_original} needs to be replaced with $\bm{\Theta}$ expressed in Eqn.~\ref{eq:large_sigma3} by taking the tensor product of $\bm{\sigma}_3$ with $\bm{I}_{2N\times 2N}$. Then, the integrand in Eqn.~\ref{eq:gen_W} becomes:
	\begin{equation}\begin{split}
	tr\left( \frac{1}{4 \pi i}
		\begin{bmatrix}
			1 & 0 && 0 & 0   \\
			0 & -1 && 0 & 0  \\
			&\vdots&\ddots&\vdots&\\
			0 & 0 && 1  & 0  \\
			0 & 0 && 0 & -1  \\
		\end{bmatrix}
		\begin{bmatrix}
			H^{-1}_{1,1} & H^{-1}_{1,2} && H^{-1}_{1,2N-1} & H^{-1}_{1,2N}   \\
			H^{-1}_{2,1} & H^{-1}_{2,2} && H^{-1}_{2,2N-1} & H^{-1}_{2,2N}  \\
			&\vdots&\ddots&\vdots&\\
			H^{-1}_{2N-1,1} & H^{-1}_{2N-1,2} && H^{-1}_{2N-1,2N-1} & H^{-1}_{2N-1,2N}   \\
			H^{-1}_{2N,1} & H^{-1}_{2N,2} && H^{-1}_{2N,2N-1} & H^{-1}_{2N,2N}  \\
		\end{bmatrix}
		\begin{bmatrix}
			0 & 0 && 0 & iac_2e^{-ika}   \\
			0 & 0 && 0 & 0  \\
			&\vdots&\ddots&\vdots&\\
			0 & 0 && 0  & 0  \\
			-iac_2e^{ika} & 0 && 0 & 0  \\
		\end{bmatrix}\right)\\
			= tr\left(\frac{a c_2}{4 \pi}
		\begin{bmatrix}
			-H^{-1}_{1,2N}e^{ika} & 0 && 0 & H^{-1}_{1,1}e^{-ika}   \\
		H^{-1}_{2,2N}e^{ika} & 0 && 0 & -H^{-1}_{2,1}e^{-ika}  \\
		&\vdots&\ddots&\vdots&\\
			-H^{-1}_{2N-1,2N}e^{ika}& 0 && 0  & H^{-1}_{2N-1,1}e^{-ika}  \\
			H^{-1}_{2N,2N}e^{ika} & 0 && 0 & -H^{-1}_{2N,1}e^{-ika}  \\
		\end{bmatrix}\right),
	\end{split}\end{equation}

where $H^{-1}_{i,j}$ denotes the element in the $i$th row and $j$th column of the inverse of $\bm{H}'$. This integrand then becomes

\begin{equation}
	Integrand = -\frac{a c_2}{4 \pi} \left(H^{-1}_{1,2N}e^{ika} + H^{-1}_{2N,1}e^{-ika}\right)
	\label{eq:S_integrand}
\end{equation}

A simple algebric derivation yields the values of $H^{-1}_{1,2N}$ and $H^{-1}_{2N,1}$, which are expressed as
\begin{eqnarray}
	 H^{-1}_{1,2N}=\frac{1}{c_2}\frac{(-1)^N}{\left(\frac{c_1}{c_2}\right)^N-(-1)^N e^{ika}},\\ \label{S_inverH12N}
	 H^{-1}_{2N,1}=\frac{1}{c_2}\frac{(-1)^N}{\left(\frac{c_1}{c_2}\right)^N-(-1)^N e^{-ika}}.\label{S_inverH2N1}
\end{eqnarray}
The $c_2$ in the front of both terms is related to the second-hierarchical-level (2nd-HL) unit cell's terminating spring.   

Substituting them into Eqn.~\ref{eq:S_integrand}, Eqn.~\ref{eq:gen_W} becomes
\begin{eqnarray} \begin{split}
	W =\int^{\pi/a}_{-\pi/a} -\frac{ac_2}{4 \pi} \left[\frac{1}{c_2}\frac{(-1)^N e^{ika}}{\left(\frac{c_1}{c_2}\right)^N-(-1)^N e^{ika}}+\frac{1}{c_2}\frac{(-1)^N e^{-ika} }{\left(\frac{c_1}{c_2}\right)^N-(-1)^N e^{-ika}}\right] dk
	&\\
	=\int^{\pi/a}_{-\pi/a}\frac{a}{4 \pi} \left[\frac{1}{1-(-1)^N\left(\frac{c_1}{c_2}\right)^N e^{-ika}}+\frac{1}{1-(-1)^N \left(\frac{c_1}{c_2}\right)^N e^{ika}}\right] dk
	&\\
	=\int^{\pi/a}_{-\pi/a}\frac{a}{4 \pi} \left[\frac{1}{1-X^-}+\frac{1}{1-X^+}\right] dk,
	\label{S_gen_W}
	\end{split}
\end{eqnarray} 
where $X^{\pm}=(-1)^N \left(\frac{c_1}{c_2}\right)^N e^{\pm ika}$ for simplicity.

When two identical first-hierarchical-level (1st-HL) domain walls separated by $d$ 1st-HL unit cells (pairs of masses) are included in a 2nd-HL unit cell containing a total of $D$ 1st-HL unit cells, we can replace $N$ with $D$. Meanwhile, the arrangement of $c_1$ and $c_2$, $i.e.$, the 1st-HL unit cell, is flipped for the masses between the two interfaces with a spacing of $d$. Thus, $X^{\pm}$ in Eqn.~\ref{S_gen_W} becomes
\begin{eqnarray} \begin{split}
	X^{\pm}=(-1)^{D}\left(\frac{c_1}{c_2}\right)^{D-d}\left(\frac{c_2}{c_1}\right)^{d}e^{\pm ika}
    &\\
    =(-1)^D \left(\frac{c_1}{c_2}\right)^{D-2d} e^{\pm ika}.
	\label{S_gen_Xpm}
    \end{split}
\end{eqnarray}

To solve for $W$, we can set $X^+=Az$ where $A=(-1)^D \left(\frac{c_1}{c_2}\right)^{D-2d}$ and $z=e^{ika}$. Then $X^-=\frac{A}{z}$. $dk$ in Eqn.~\ref{S_gen_W} can then be expressed as 

\begin{equation}
	dk=-\frac{i}{ae^{ika}}dz=-\frac{i}{az}dz.
	\label{S_dk}
\end{equation}

Substituting the expression of $z$ and $dk$ above into Eqn.~\ref{S_gen_W}, we get:

\begin{equation}\begin{split}
	W=\frac{a}{4 \pi}\oint_{c} \left(\frac{z}{z-A}+\frac{1}{1-Az}\right)\left(-\frac{i}{az}\right) dz
	&\\
	=-\frac{i}{4\pi}\left[\oint_c \frac{1}{z-A}dz +\oint_c\frac{1}{(1-Az)z}dz\right],
		\label{S_W_contour}
	\end{split}
\end{equation}
where $C$ is the closed contour with $\abs{z}=1$ winding from $-\pi$ to $\pi$ in the counter-clockwise direction. Using the residue theorem, we can obtain the solutions of Eqn.~\ref{S_W_contour} by comparing the poles determined by $\abs{A}$ with $\abs{z}$, which gives:
\begin{eqnarray}
	W=
	\begin{cases}
		-\frac{i}{4\pi}\cdot{2\pi i}\left[0+0\right]=0  \quad\text{for} \, &\abs{A}>1\\
		-\frac{i}{4\pi}\cdot{2\pi i}\left[1+1\right]=1  \quad\text{for} \, &\abs{A}<1
	\end{cases}
	\label{S_W_solution}
\end{eqnarray}

Since whether the poles are located in the positive or negative $x-$axis does not affect the winding number solution, the only term in the expression of $A$ that determines whether the pole is located within $\abs{z}=1$ is $\abs{A}=\left(\frac{c_1}{c_2}\right)^{D-2d}$, which we will later refer to as the \textit{conditioning coefficient}. Thus, solutions of $W$ are reduced to:
\begin{eqnarray}
	W=
	\begin{cases}
		0  \quad\text{for} \, &\left(\frac{c_1}{c_2}\right)^{D-2d}>1\\
		1  \quad\text{for} \, &\left(\frac{c_1}{c_2}\right)^{D-2d}<1
	\end{cases}
	\label{S_W_solution_final}
\end{eqnarray}
This final expression of the winding number $W$ suggests its dependence on both the spring arrangement near the boundaries of the 2nd-HL unit cell, $i.e.$, the relative values of $c_1$ and $c_2$, and the relative distance between the two 1st-HL domain walls, $d$, with respect to the size of the 2nd-HL cell size, $D$.
\section{Topology of a generalized one-dimensional higher-hierarchical-level unit cell}

Before extending the above analysis to an $n$th-HL unit cell, we revisit the derivation of the conditioning coefficient $\abs{A}$ in the expression of the 2nd-HL winding number (Eqn.~\ref{S_W_solution}), where $\abs{A}=\left(\frac{c_1}{c_2}\right)^{D-2d}$. The exponent of $\frac{c_1}{c_2}$ arises from the arrangement of the 1st-HL springs between and outside of the two interfaces in the 2nd-HL unit cell. To streamline the discussion in this section, we relabel all quantities with subscripts indicating the hierarchical level, yielding:
\begin{eqnarray}
	\abs{A}_2=\left(\frac{c_2}{c_1}\right)^{d_2}\left(\frac{c_1}{c_2}\right)^{D_2-d_2}\\
    =\left(\frac{c_1}{c_2}\right)^{D_2-2d_2},
	\label{A2_re}
\end{eqnarray}

For the 1st-HL unit cell - a simple pair of masses and springs, the conditioning coefficient is given by $\abs{A}_1=\frac{c_1}{c_2}$, such that:
\begin{eqnarray}
	W_1=
	\begin{cases}
		0  \quad\text{for} \, &\abs{A}_1>1\\
		1  \quad\text{for} \, &\abs{A}_1<1,
	\end{cases}
	\label{S_W_A1}
\end{eqnarray}
In the 2nd-HL unit cell, the spring arrangements inside and outside the two 1st-HL domain walls are inverted. Denoting the conditioning coefficient of the 1st-HL unit cell terminating the 2nd-HL unit cell as $\abs{A}_1^{out}=\frac{c_1}{c_2}$, then that between the domain walls becomes $\abs{A}_1^{in}=\frac{c_2}{c_1}$. This leads the expression of $\abs{A}_2$ to:
\begin{eqnarray}
\abs{A}_2=\left(\abs{A}_1^{in}\right)^{d_2}\left(\abs{A}_1^{out}\right)^{D_2-d_2}.
	\label{A2_re2}
\end{eqnarray}

In the construction of 3rd-HL unit cells, the geometric parameters $D_2$ (2nd-HL unit cell size) and $d_2$ (number of 1st-HL cells between the two 1st-HL domain walls) may differ between the interior and exterior regions of the 2nd-HL domain walls. We therefore distinguish between interior parameters $D_2^{in}$ and $d_2^{in}$ between the domain walls, and exterior parameters $D_2^{out}$ and $d_2^{out}$ outside the domain walls. This spatial variation is naturally reflected in the 2nd-HL conditioning coefficients, which take the forms, $\abs{A}_2^{in}=\left(\frac{c_1}{c_2}\right)^{D_2^{in}-2d_2^{in}}$ and $\abs{A}_2^{out}=\left(\frac{c_1}{c_2}\right)^{D_2^{out}-2d_2^{out}}$, for the interior and exterior regions, respectively. 

Let $D_3$ denote the total number of 2nd-HL unit cells, with $d_3$ representing the number of these cells located between the 2nd-HL domain walls. Under the assumption that all terminating 1st-HL unit cells maintain identical arrangements across the 2nd-HL cells, we can express the count of flipped 1st-HL unit cells ($N_{fl}$) where the $c_1$ and $c_2$ arrangments are inverted as $N_{fl}=d_3d_2^{in}+(D_3-d_3)d_2^{out}$, which accounts for all 1st-HL cells between the two 1st-HL domain walls in all 2nd-HL unit cells. The number of unflipped 1st-HL unit cells ($N_{uf}$), where spring arrangements are preserved as the terminating boundaries, becomes $N_{uf}=d_3\left(D_2^{in}-d_2^{in}\right)+(D_3-d_3)\left(D_2^{out}-d_2^{out}\right)$. Substituting these expressions into the formulation of  $\abs{A}_3$ yields:
\begin{eqnarray}
	\abs{A}_3=\left(\frac{c_1}{c_2}\right)^{N_{uf}}\left(\frac{c_2}{c_1}\right)^{N_{fl}}\\
    =\left(\frac{c_1}{c_2}\right)^{d_3\left(D_2^{in}-d_2^{in}\right)+\left(D_3-d_3\right)\left(D_2^{out}-d_2^{out}\right)}\left(\frac{c_2}{c_1}\right)^{d_3d_2^{in}+\left(D_3-d_3\right)d_2^{out}}\\
    =\left(\frac{c_1}{c_2}\right)^{d_3\left(D_2^{in}-2d_2^{in}\right)}\left(\frac{c_1}{c_2}\right)^{\left(D_3-d_3\right)\left(D_2^{out}-2d_2^{out}\right)}\\
    =\left(\abs{A}_2^{in}\right)^{d_3}\left(\abs{A}_2^{out}\right)^{D_3-d_3}.
	\label{S_gen_Xpm_3rd_1}
\end{eqnarray}

When the spring configurations terminating the 2nd-HL unit cells differ between those inside and outside the 2nd-HL domain walls (with opposite $c_1$ and $c_2$ arrangements), the counts of flipped and unflipped 1st-HL cells are respectively given by $N_{fl}=d_3\left(D_2^{in}-d_2^{in}\right)+(D_3-d_3)d_2^{out}$ and $N_{uf}=d_3d_2^{in}+(D_3-d_3)\left(D_2^{out}-d_2^{out}\right)$. With the interior 2nd-HL conditioning coefficient expressed as $\abs{A}_2^{in}=\left(\frac{c_2}{c_1}\right)^{D_2^{in}-2d_2^{in}}$, the resulting 3rd-HL conditioning coefficient takes the form:
\begin{eqnarray}
	\abs{A}_3=\left(\frac{c_1}{c_2}\right)^{N_{uf}}\left(\frac{c_2}{c_1}\right)^{N_{fl}}\\
    =\left(\frac{c_1}{c_2}\right)^{d_3d_2^{in}+(D_3-d_3)\left(D_2^{out}-d_2^{out}\right)}\left(\frac{c_2}{c_1}\right)^{d_3\left(D_2^{in}-d_2^{in}\right)+(D_3-d_3)d_2^{out}}\\
    =\left(\frac{c_2}{c_1}\right)^{d_3\left(D_2^{in}-2d_2^{in}\right)}\left(\frac{c_1}{c_2}\right)^{\left(D_3-d_3\right)\left(D_2^{out}-2d_2^{out}\right)}\\
    =\left(\abs{A}_2^{in}\right)^{d_3}\left(\abs{A}_2^{out}\right)^{D_3-d_3}.
	\label{S_gen_Xpm_3rd_2}
\end{eqnarray}

The preceding derivation reveals that the conditioning coefficient for the $(n-1)$th HL can be expressed through its exponential index $Nd_{n-1}$ as:
\begin{eqnarray}
    \abs{A}_{n-1}=\left(\frac{c_1}{c_2}\right)^{Nd_{n-1}},
\end{eqnarray}
where $Nd_i$ represents the net difference between unflipped and flipped springs in the $i$th-HL unit cell. This formulation naturally extends to the $n$th HL through continued accumulation of these differences.

Crucially, the distinct ($n-1$)th-HL configurations inside ($\abs{A}_{n-1}^{in}$) and outside ($\abs{A}_{n-1}^{out}$) the domain walls - a necessary condition for domain wall formation - guarentee different net differences $Nd_{n-1}^{in}$ and $Nd_{n-1}^{out}$, respectively. Consequently, the cumulative difference $Nd_n$ at the $n$th HL becomes:
\begin{eqnarray}
    Nd_n=\left(D_n-d_n\right)Nd_{n-1}^{out}+d_nNd_{n-1}^{in}.
\end{eqnarray}
The $n$th-HL conditioning coefficient then becomes
\begin{eqnarray}
    \abs{A}_n=\left(\frac{c_1}{c_2}\right)^{Nd_d}\\
    =\left(\frac{c_1}{c_2}\right)^{\left(D_n-d_n\right)Nd_{n-1}^{out}+d_nNd_{n-1}^{in}}\\
    =\left[\left(\frac{c_1}{c_2}\right)^{Nd_{n-1}^{in}}\right]^{d_n}\left[\left(\frac{c_1}{c_2}\right)^{Nd_{n-1}^{out}}\right]^{D_n-d_n}\\
    =\left(\abs{A}_{n-1}^{in}\right)^{d_n}\left(\abs{A}_{n-1}^{out}\right)^{D_n-d_n}.
\end{eqnarray}

 The above derivation demonstrates that the current conditioning coefficient, $\abs{A}_n$ depends on three key factors: 
 \begin{itemize}
	\item The number of $(n-1)$th-HL cells between the two $(n-1)$th-HL domain walls ($d_n$).
	\item The total count of $(n-1)$th-HL cells in the $n$th-HL unit cell ($D_n$).
	\item The hierarchical conditioning coefficients from the previous level, namely the interior ($\abs{A}_{n-1}^{in}$) and exterior ($\abs{A}_{n-1}^{out}$) coefficients relative to the $(n-1)$-HL domain walls.
\end{itemize}

The winding number at the $n$-HL is then determined by 
\begin{equation}
	 	W_n = 
\begin{cases}
	1 \quad \text{for} \, & \abs{A}_n<1 \\
	0 \quad \text{for} \, & \abs{A}_n>1.
\end{cases}
\end{equation}

\begin{figure}[!htbp]
	\centering
	\includegraphics[scale=0.48]{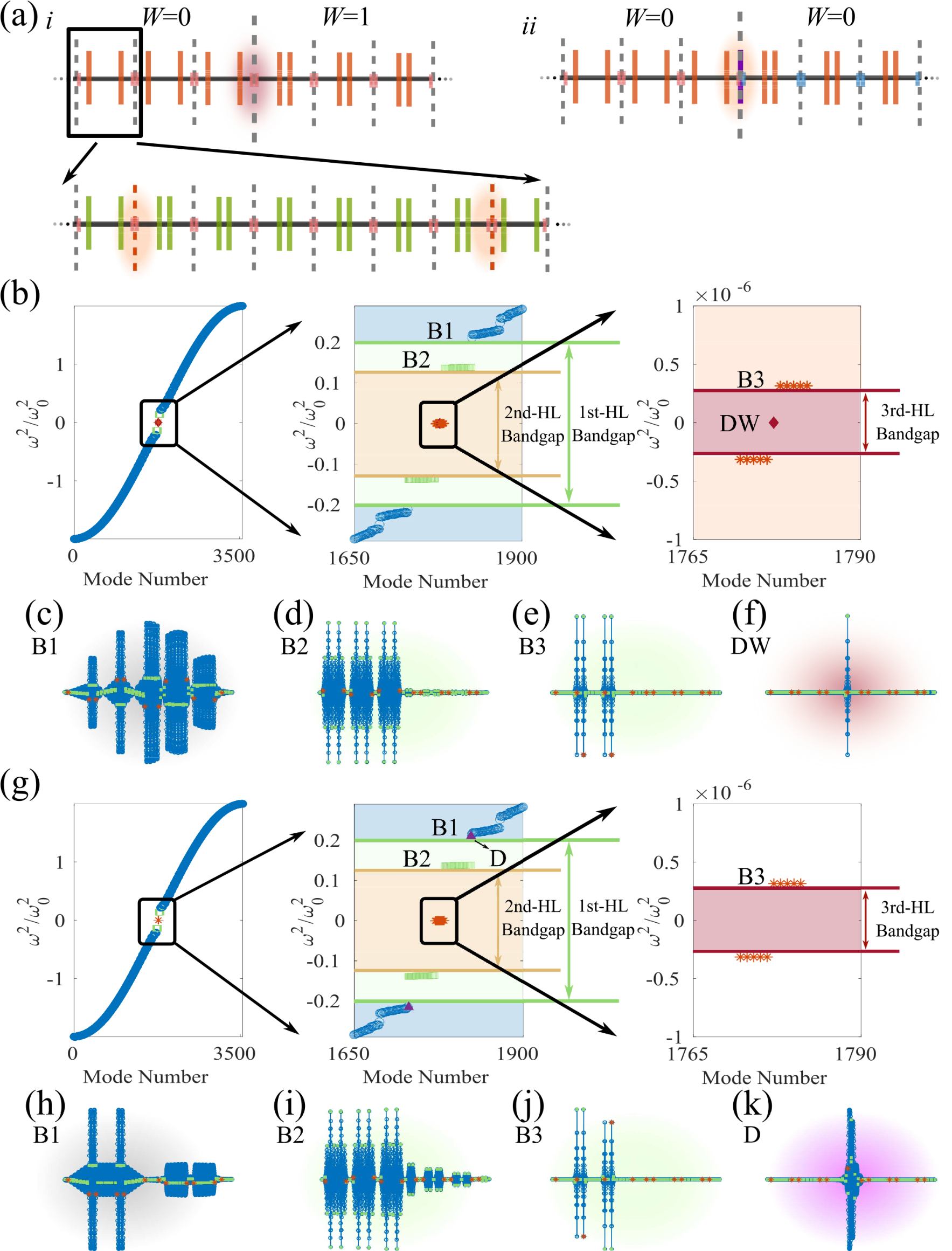} 
	\caption{Spectral and modal characteristics of one-dimensional third-hierarchical-level (3rd-HL) domain wall states. (a) Schematics of two finite
lattices featuring 3rd-HL domain walls, each with six 3rd-HL cells per side. (b,g) Normalized eigenfrequencies ($\omega^2/\omega_0^2$) for both scenarios in (a), with zoomed-in views highlighting: 1st-HL bulk modes (B1, blue), 1st-HL bandgaps, (green), 2nd-HL bulk modes (B2, green), 2nd-HL bandgaps (orange), 3rd-HL bulk modes (B3, red stars), and 3rd-HL bandgap (red) hosting 3rd-HL domain wall (red DW, diamond)orange. The defect mode in (g) is marked as a purple triangle. (c-f,h-k) Representative mode shapes of B1, B2, B3, DW, and D (if present). Green squares and red stars mark 1st- and 2nd-HL domain walls, respectively.}
	\label{fig:3rd_HL}
\end{figure}

Two illustrative cases demonstrating how a higher-HL winding number ($W_n$) predicts the number of TDWSs at the domain wall formed by different types of 3rd-HL lattices are presented in Fig.~\ref{fig:3rd_HL}. When $\Delta W_3=1$, a zero-frequency TDWS emerges within the newly opened bandgap created by repositioning the 2nd-HL domain walls inside each 3rd-HL unit cell. In contrast, if the 2nd-HL domain-wall positions remain the same but spring constants $c_1$ and $c_2$ are swapped on the right side, the winding number becomes $W_3=0$ there. This configuration eliminates the TDWS while keeping the 3rd-HL bandgap open, instead producing a trivial defect mode (D) localized near near the 3rd-HL domain wall within the 1st-HL bandgap. 

\section{Participation Ratio}
To distinguish between 1st-HL bulk modes (B1), 2nd-HL bulk modes (B2) within the 1st-HL bandgap, and the 2nd-HL TDWS (DW), we evaluate the participation ratio ($PR$) for each eigenmode $m$ in a finite system:
\begin{equation}
	PR(m) = \frac{(\Sigma_i |u_i(m)|^2)^2}{\Sigma_i |u_i(m)|^4}
\end{equation} 
where $u_i(m)$ represents the displacement of the $i$th mass of the $m$th eigenmode. For a system with M masses, $\frac{PR}{M} \rightarrow 1$ indicates uniform displacement across all masses, implying a bulk mode. However, while bulk modes typically exhibit higher $\frac{PR}{M}$ values, the defining characteristic is their convergence to a constant number as $M$ increases, reflecting their insensitivity to system size. In contrast, localized states, such as TDWSs, exhibit $\frac{PR}{M} \rightarrow 0$ with increasing $M$, as their spatial confinement prevents participation from distant masses. 

\begin{figure}[!htbp]
	\centering
	\includegraphics[scale=0.27]{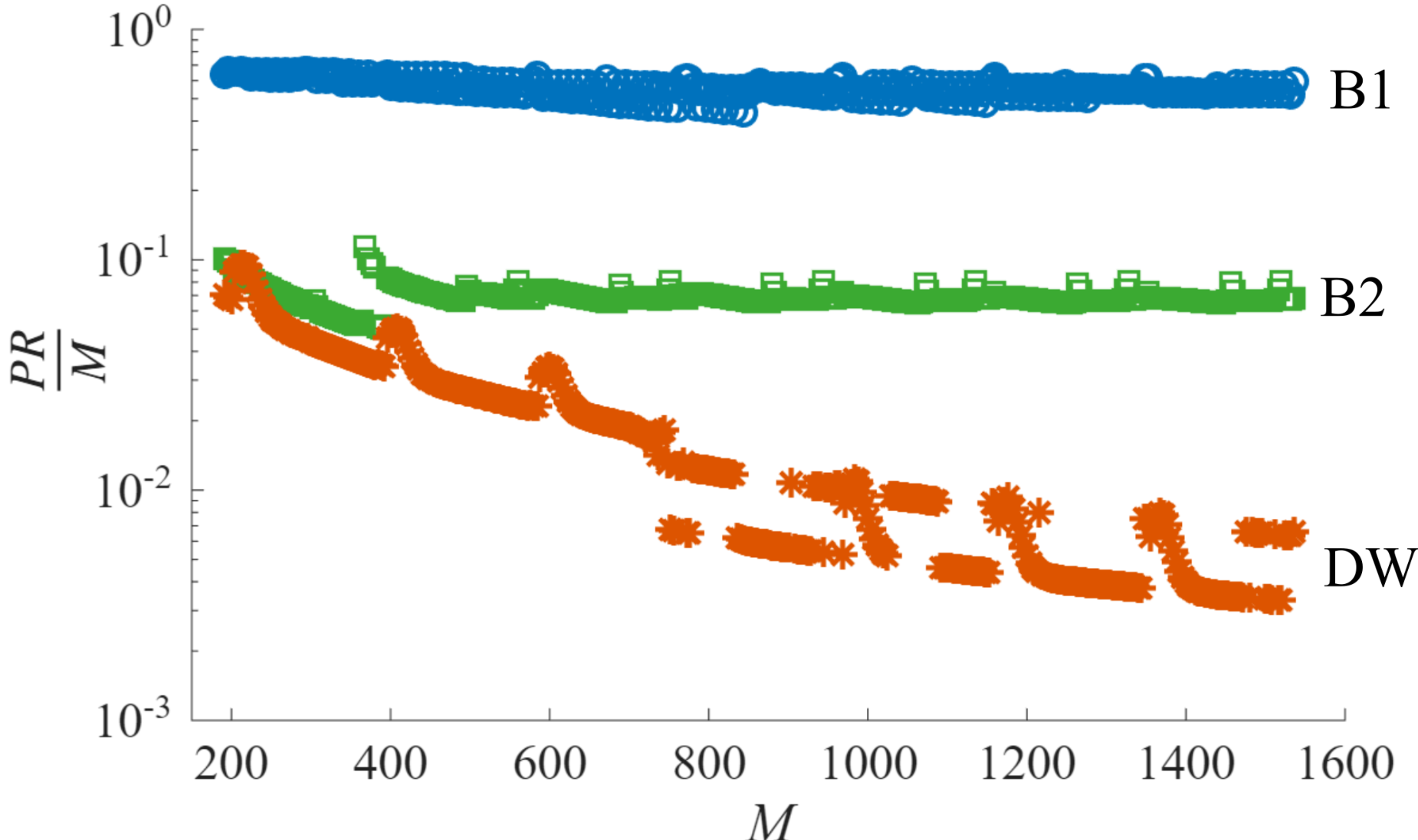} 
	\caption{Normalized participation ratio ($\frac{PR}{M}$) as a function of system size ($M$) for the 1st-HL bulk modes (B1, blue circles), 2nd-HL bulk modes (B2, green squares), and 2nd-HL TDWS (DW, orange stars), corresponding to the modes shown in Fig.~\ref{fig:interfacing_plot} (b-f). The asymptotic behavior demonstrates the extended nature of bulk modes (B1/B2, converging to constants) versus the localized character of the TDWS (DW, decaying to zero).}
	\label{fig:PR}
\end{figure}

Fig.~\ref{fig:PR} shows $\frac{PR}{M}$ calculations for Scenario $\it{i}$ in Fig.~\ref{fig:interfacing_plot}. The B1 and B2 modes in Fig.~\ref{fig:interfacing_plot}(c-e) both demonstrate bulk behavior, with $\frac{PR}{M}$ converging to constant values as $M$ grows. However, B1 modes exhibit significantly higher $\frac{PR}{M}$ values than B2 modes. This distinction arises because B2 modes, while delocalized across the entire lattice, are composed of coupled 1st-HL TDWSs, which are localized excitations within individual 2nd-HL cells that collectively form extended bulk modes. Consequently, B2 modes retain weaker participation than B1 modes, yet both converge to size-independent values. In contrast, the DW mode (2nd-HL TDWS) shows $\frac{PR}{M} \rightarrow 0$ asymptotically, confirming its localization at the 2nd-HL domain wall and insensitivity to system size.  

\section{Two-Dimensional Su-Schrieffer-Heeger Model}

The two-dimensional (2D) mechanical SSH model examined in this work consists of identical masses connected by springs with alternating stiffnesses, as illustrated in Fig.~\ref{fig:2dunit_supp}(a,b). The equations of motion for a 1st-HL unit cell can be derived via Newton's Second Law: 

\begin{figure*}[!htbp]
	\centering
	\includegraphics[scale=0.35]{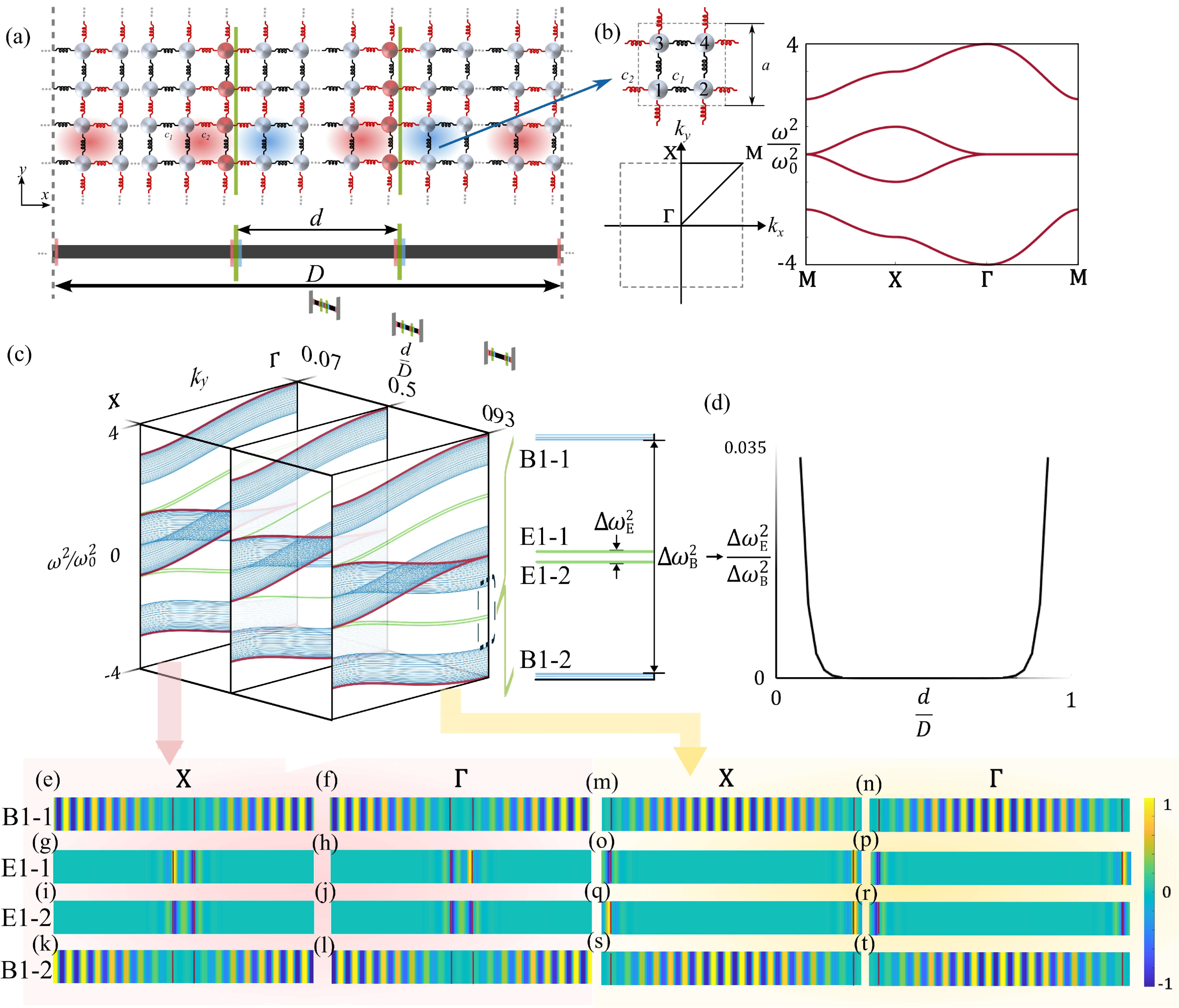} 
	\caption{(a)Schematic of the 2nd-HL lattice comprising 74 masses per row along the $x$-axis connected by springs with alternating stiffness ($c_1$, black; $c_2$, red). Red and blue shading highlights the two simulated rows under Bloch boundary conditions in both $x$- and $y$-directions. Vertical green bars mark two 1st-HL domain walls. Parameters $d$ and $D$ denote the number of mass pairs in each row between 1st-HL domain walls and the total number within the 2nd-HL unit cell, respectively. (b) Zoomed-in view of the 1st-HL unit cell (top left) with mass sequence in the Hamiltonian matrix labeled, its reciprocal space and irreducible Brillouin zone (bottom left), and corresponding phonon dispersion (right). (c) Full phonon dispersion of the 2nd-HL system for $c_1=1.5$ and $c_2=0.5$ along $k_y$ (denoted X-$\mathrm\Gamma$) showing bulk bands (B1-1/2, blue), topological domain-wall states (TDWSs, E1-1/2 green), and 1st-HL bulk dispersions (red) for varying $d/D$ ratios. (d) Normalized bandgap ratio $\Delta \omega_E^2/\Delta \omega_B^2$ at $\mathrm{\Gamma}$ between the 1st-HL TDWSs (E1-1/2) and bulk bands (B1-1/2) as a function of $d/D$. (e-t) Mode shapes at $X$ and $\mathrm{\Gamma}$ for bands B1-1/2 and E1-1/2, comparing configurations with (e-l) $d/D=0.07$ and (m-t) $d/D=0.93$. Red bars represent the two 1st-HL domain walls within each 2nd-HL cell.}
	\label{fig:2dunit_supp}
\end{figure*}

\begin{eqnarray}
	m\ddot{u}_1^{m,n} = c_1(u_2^{m,n} - u_1^{m,n}) - c_2(u_1^{m,n} - u_2^{m-1,n})+c_1(u_3^{m,n}-u_1^{m,n})-c_2(u_1^{m,n}-u_3^{m,n-1}), \nonumber \\
	m\ddot{u}_2^{m,n} = c_2(u_1^{m+1,n} - u_2^{m,n}) - c_1(u_2^{m,n} - u_1^{m,n})+c_1(u_4^{m,n}-u_2^{m,n})-c_2(u_2^{m,n}-u_4^{m,n-1}),\nonumber\\
    m\ddot{u}_3^{m,n}= c_1(u_4^{m,n} - u_3^{m,n}) - c_2(u_3^{m,n} - u_4^{m-1,n})+c_2(u_1^{m,n+1}-u_3^{m,n})-c_1(u_3^{m,n}-u_1^{m,n}), \nonumber \\
    m\ddot{u}_4^{m,n}= c_2(u_3^{m+1,n} - u_4^{m,n}) - c_1(u_4^{m,n} - u_3^{m,n})+c_2(u_2^{m,n+1}-u_4^{m,n})-c_1(u_4^{m,n}-u_2^{m,n}), \nonumber \\
	\label{eq:2dspring_mass_gov}
\end{eqnarray}
where $u_i^{m,n}$ denotes the out-of-plane displacement of the $i$th mass in the $(m,n)$th unit cell along the $x$ and $y$ axes. We assume small displacements such that spring forces act linearly. 

Applying the plane-wave ansatz with Bloch periodic boundary conditions, the displacement vector at time $t$ is expressed as:

\begin{eqnarray}
	{\mathbfit u}^{m,n}(t)={\tilde{\mathbfit{u}}}(\mathbfit{k})e^{i(\mathbfit{k\cdot r}_{m,n}-\omega t)},
	\label{eq:2d_planewave}
\end{eqnarray}
where ${\mathbfit u}^{m,n}=[u_1^{m,n},u_2^{m,n},u_3^{m,n},u_4^{m,n}]^\mathrm{T}$, $\mathbfit{k}=(k_x,k_y)$ is the wavevector, and $\mathbfit{r}_{m,n}=(ma,na)$ is the position vector of the $(m,n)$th unit cell, with $a$ denotes the lattice constant in both $x$- and $y$-directions. 

Substituting Eqn.~\ref{eq:2d_planewave} into the equations of motion Eqns.~\ref{eq:2dspring_mass_gov} yields the eigenvalue problem:
\begin{eqnarray}
	[\bm{H}(\mathbfit{k})-\omega^2m]{\tilde{\mathbfit{u}}}(\mathbfit{k})=0,
	\label{2d_goveqn}
\end{eqnarray}
where $\bm{H}(\mathbfit{k})$ is the stiffness matrix of the periodic system:
\begin{equation}
	\bm{H}(\mathbfit{k}) = 
	\begin{bmatrix}
		2(c_1+c_2) & -c_1 - c_2e^{-ik_xa} & -c_1-c_2e^{-ik_ya} & 0 \\
		-c_1 - c_2e^{ik_xa} & 2(c_1+c_2) & 0 & -c_1-c_2e^{-ik_ya} \\
        -c_1-c_2e^{ik_ya} & 0 & 2(c_1+c_2) & -c_1-c_2e^{-ik_xa} \\
        0 & -c_1-c_2e^{ik_ya} & -c_1-c_2e^{ik_xa} & 2(c_1+c_2)
	\end{bmatrix}.
	\label{eq:2d_Hmatrix}
\end{equation}
To isolate the topological features, we subtract the uniform diagonal term to obtain a chiral matrix:
\begin{eqnarray}
	\bm{H}'(k)=\bm{H}(k)-2(c_1+c_2)\bm{I},
	\label{eq:2d_chiral}
\end{eqnarray}
where $\bm{I}$ is the identity matrix.In the tight-binding analogy, $c_1$ and $c_2$ represent intra- and inter-cell hopping parameters, respectively. Using $h_{12}=-c_1-c_2e^{ik_xa}$, $h_{13}=-c_1-c_2e^{ik_ya}$, $h_{24}=-c_1-c_2e^{ik_ya}$, $h_{34}=-c_1-c_2e^{ik_xa}$, we recover Eqn.~\ref{eq:2D_Hmatrix} in the main text.

Solving for the eigenvalues of $\bm{H}'(\mathbfit{k})$ yields a band structure symmetric about the zero-frequency (or zero-energy, in quantum systems) state, $i.e.$, $\frac{\omega^2}{\omega_0^2} = 0$, where \( \omega_0^2 = \frac{2(c_1 + c_2)}{m} \), as shown in Fig. \ref{fig:2dunit_supp} (b). Notably, when $c_1\neq c_2$, no bandgap opens at $\frac{\omega^2}{\omega_0^2} = 0$; instead, an indirect bandgap emerges at $\frac{\omega^2}{\omega_0^2} = \pm 2$, hosting topological states, as illustrated in Fig.~\ref{fig:2dunit_supp} (c) and Fig.~\ref{fig:2D_Interface_distance} (c) in the main text.

Expanding the unit cell along the $x$-direction to include a total of $D$ mass pairs in each row with two TDWSs separated by $d$ mass pairs, $\bm{H}'(\mathbfit{k}$ can be generalized accordingly:

\begin{equation}
	\bm{H}'(\mathbfit{k}) = 
	\begin{bmatrix}
		0 & -c_1 & 0 & \dots & -c_2e^{-ik_xa} & -c_1-c_2e^{-ik_ya} & 0 & \dots & 0 \\
		-c_1 &0&-c_2 && 0 &0 &-c_1-c_2e^{-ik_ya} &\dots&0\\
         &&&\vdots&&&\vdots&\\
        \dots&-c_2 &0&-c_2&\dots&-c_2&\dots&-c_1-c_2e^{-ik_ya} &\dots&\\
        &&&\vdots&&&\vdots&\\
        &\dots&-c_1-c_2e^{ik_ya}&\dots&0&-c_2&\dots&-c_2&\dots\\
        &&&\vdots&&&\vdots&\\
        &&\dots&-c_1-c_2e^{ik_ya}&&\dots&-c_2&0&-c_1\\
        &&\dots&&-c_1-c_2e^{ik_ya}&-c_2e^{ik_xa}&\dots&-c_1&0
	\end{bmatrix}_{2D\times 2D}.
	\label{eq:2d_bigHmatrix}
\end{equation}

By computing the eigenvalues along the $k_x$ and $k_y$ directions, we obtain the dispersion relations and corresponding mode shapes shown in Fig.~\ref{fig:2dunit_supp}(c–t) and Fig.~\ref{fig:2D_Interface_distance}(c–t). Notably, while mode shapes along $k_x$ exhibit parity inversion between the X and $\Gamma$ points, no such inversion occurs along $k_y$. This directional asymmetry confirms that the system’s topological characteristics are separable in the two dimensions, validating our approach of analyzing topology along a single direction without interference from the other.

\end{document}